\documentclass[fleqn,usenatbib]{mnras}

\usepackage[T1]{fontenc}

\DeclareRobustCommand{\VAN}[3]{#2}
\let\VANthebibliography\thebibliography
\def\thebibliography{\DeclareRobustCommand{\VAN}[3]{##3}\VANthebibliography}

\usepackage{graphicx}	%
\usepackage{amsmath}	%
\usepackage{amssymb}	%
\usepackage[dvipsnames]{xcolor} 
\usepackage{newtxtext,newtxmath}
\usepackage[normalem]{ulem}

\newcommand{\LIGOlabMIT}{LIGO Laboratory, Massachusetts Institute of Technology, 185 Albany St, Cambridge, MA 02139, USA}
\newcommand{\MKI}{Department of Physics and Kavli Institute for Astrophysics and Space Research, Massachusetts Institute of Technology, \\ 77 Massachusetts Ave, Cambridge, MA 02139, USA}

\title[BNS mass distribution systematic error]{The effect of spin mismodeling on gravitational-wave measurements of the binary neutron star mass distribution}

\author[Biscoveanu et al.]{
Sylvia Biscoveanu,$^{1,2}$\thanks{sbisco@mit.edu}
Colm Talbot,$^{1,2}$
Salvatore Vitale$^{1,2}$
\\
$^{1}$\LIGOlabMIT\\
$^{2}$\MKI
}

\date{Accepted 2022 February 4. Received 2022 February 1; in original form 2021 November 26}

\pubyear{2021}

\begin{document}
\label{firstpage}
\pagerange{\pageref{firstpage}--\pageref{lastpage}}
\maketitle

\begin{abstract}
The binary neutron star (BNS) mass distribution measured with gravitational-wave observations has the potential to reveal information about the dense matter equation of state, supernova physics, the expansion rate of the universe, and tests of General Relativity. As most current gravitational-wave analyses measuring the BNS mass distribution do not simultaneously fit the spin distribution, the implied population-level spin distribution is the same as the spin prior applied when analyzing individual sources. In this work, we demonstrate that introducing a mismatch between the implied and true BNS spin distributions can lead to biases in the inferred mass distribution. This is due to the strong correlations between the measurements of the mass ratio and spin components aligned with the orbital angular momentum for individual sources. We find that applying a low-spin prior which excludes the true spin magnitudes of some sources in the population leads to significantly overestimating the maximum neutron star mass and underestimating the minimum neutron star mass at the population level with as few as six BNS detections. The safest choice of spin prior that does not lead to biases in the inferred mass distribution is one which allows for high spin magnitudes and tilts misaligned with the orbital angular momentum.
\end{abstract}


\section{Introduction}

The growing catalog of compact binary mergers detected in gravitational waves~\citep{LIGOScientific:2021djp} has provided a novel means to probe the properties of black holes and neutron stars~\citep{LIGOScientific:2021psn, Wong:2020ise, Zevin:2020gbd, Roulet:2021hcu, Bouffanais:2021wcr}. The neutron star mass distribution has the potential to independently yield information on the dense matter equation of state (EoS) via the maximum mass, $M_{\mathrm{TOV}}$, beyond which the internal pressure of the neutron star can no longer support it against gravitational collapse to a black hole~\citep{Miller:2019nzo, Landry:2020vaw, Chatziioannou:2020msi, Legred:2021hdx}. %
The value of the maximum mass depends on the yet-unknown EoS and the neutron star spin~\citep{Lasota:1995eu}, although astrophysical processes may prevent the formation of neutron stars with mass up to $M_{\mathrm{TOV}}$ in some scenarios. The mass distribution can also be used to constrain the supernova physics leading to neutron star formation~\citep{Pejcha:2012at, Vigna-Gomez:2018dza}, the astrophysical stochastic gravitational-wave background~\citep{Zhu:2012xw, LIGOScientific:2017zlf}, the rate of expansion of the Universe~\citep{Chernoff:1993th, Finn:1995ah, Taylor:2011fs, Taylor:2012db} and alternative theories of gravity beyond General Relativity~\citep{Finke:2021eio}.

Initial measurements of the mass distribution of Galactic neutron stars detected electromagnetically suggested that their component masses follow a narrow Gaussian distribution~\citep{Thorsett:1998uc}, particularly for those found in binary systems with another neutron star~\citep{Ozel:2012ax, Kiziltan:2013oja}. More recent measurements reveal that the overall distribution of neutron star masses including those found in binary systems with other types of companions is broader~\citep{Kiziltan:2013oja, Ozel:2016oaf} and better described by a double Gaussian~\citep{Alsing:2017bbc, Antoniadis:2016hxz, Tauris:2017omb, Shao:2020bzt}. Including the first binary neutron star system observed in gravitational waves, GW170817~\citep{LIGOScientific:2017vwq}, \citet{Farrow:2019xnc} also find weak evidence for bimodality in the mass distribution of the double neutron star population alone.

Gravitational-wave observations of compact binary mergers involving at least one neutron star offer a complementary means to probe the neutron star mass distribution at extragalactic distances. While only two binary neutron star (BNS) systems have been detected in gravitational waves, these observations already suggest that there may be a distinction between the Galactic population accessible as pulsars and the gravitational-wave population~\citep{Pankow:2018iab, Galaudage:2020zst, Safarzadeh:2020efa, LIGOScientific:2021psn}. GW190425 represents the most massive BNS system ever detected, with a total mass five standard deviations away from the mean of the Galactic population~\citep{LIGOScientific:2020aai}. When analyzing the full population of neutron stars detected in gravitational waves---including neutron star-black hole mergers~\citep{LIGOScientific:2021qlt}---\citet{Landry:2021hvl} find that the masses are consistent with being uniformly distributed, although with significantly more support for high neutron star masses compared to the Galactic population. Thus, accurate and precise measurements of the neutron star mass distribution from gravitational-wave observations have the potential to elucidate the formation channels of these systems.

Such measurements utilize the framework of hierarchical Bayesian inference (e.g. \citealt{Mandel:2009pc, Thrane:2018qnx}), where the properties of the population as a whole are determined while taking into account the uncertainty on the parameters of individual sources and the selection effect introduced by the varying sensitivity of the detector to sources with different properties~\citep{Loredo_2004, Mandel:2018mve, Vitale:2020aaz}. This requires both unbiased parameter estimates for individual sources and physically realistic models for the population and detector sensitivity. One potential source of systematic error is the correlation between measurements of the intrinsic parameters describing binary neutron star systems, including the masses, spins, and tidal deformabilities of the components. \citet{Wysocki:2020myz} and \citet{Golomb:2021tll} recently demonstrated that unphysical assumptions about the equation of state (or tidal deformability) when measuring the mass distribution independently can lead to biases with as few as 37 events, emphasizing the importance of fitting these distributions simultaneously.

The mass ratio and spin components aligned with the orbital angular momentum are particularly correlated for individual sources~\citep{Cutler:1994ys, Hannam:2013uu, Berry:2014jja, Farr:2015lna, Ng:2018neg}. This is due to the fact that for a given chirp mass, $\mathcal{M} \equiv (m_{1}m_{2})^{3/5}/(m_{1} + m_{2})^{1/5}$, binaries with larger spins aligned with the orbital angular momentum will merge more slowly~\citep{Campanelli:2006uy}, while binaries with more unequal mass ratios will merge more quickly, introducing a degenerate effect on the waveform. Gravitational-wave analyses of individual BNS sources typically assume two different priors for the spin distributions, a ``low-spin'' prior and a ``high-spin'' prior~\citep{LIGOScientific:2017vwq, LIGOScientific:2020aai}. The ``low-spin'' prior restricts the maximum dimensionless spin magnitude---defined as $|\boldsymbol{\chi}| = c|\mathbf{S}|/(Gm^{2})$, where $\boldsymbol{S}$ is the spin vector of the neutron star with mass $m$---to $|\boldsymbol{\chi}| \leq 0.05$, informed by the spins of Galactic double neutron stars that will merge within a Hubble time~\citep{Lorimer:2008se, Lo:2010bj, 2018PhRvD..98d3002Z}. The ``high-spin'' prior extends to $|\boldsymbol{\chi}| \lesssim 0.89$, as allowed by available waveform models without making any assumptions about the consistency of the system with the observed Galactic population. It is worth noting that pulsars have been observed with spins as high as $|\boldsymbol{\chi}| \lesssim 0.4$, even in binary systems~\citep{Hessels:2006ze}.

In this work, we demonstrate that mismodeling the spin distribution of BNSs can lead to a bias in the recovered mass distribution. We find that for a population of sources with moderate aligned spins extending out to $|\boldsymbol{\chi}| \leq 0.4$, using individual-event mass estimates obtained with the low-spin prior without simultaneously fitting the spin distribution hierarchically leads to significant bias in both the inferred mass ratio distribution and maximum mass. Conversely, in the case of a low  aligned-spin population with $|\boldsymbol{\chi}| \leq 0.05$, the inferred mass ratio distribution under the high-spin prior is also biased, although this effect can be ameliorated by allowing for the spins to be misaligned with the orbital angular momentum. In Section~\ref{sec:methods} we describe our methodology and simulated BNS populations. In Section~\ref{sec:results} we present the results of our hierarchical inference and conclude with a discussion of the implications of our findings in Section~\ref{sec:conclusions}. We also include a demonstration of the importance of obtaining unbiased inferences for individual sources via the fallibility of the commonly-used ``probability-probability plot'' as a test of sampler performance in Appendix~\ref{ap:pp}.

\section{Methods}
\label{sec:methods}
In addition to the component masses, spins, and tidal deformabilities, quasi-circular BNS mergers are characterized by seven extrinsic parameters including the distance, sky location, time of coalescence, and inclination angle between the orbital angular momentum and observer line-of-sight. We simulate two populations of BNS systems with distinct spin distributions. Both have spins aligned with the orbital angular momentum drawn from the implied distribution on $\chi_{z}$ assuming that the magnitudes are distributed uniformly on $[0, \chi_{\max}]$ and the directions are isotropic. The first population allows for medium spins consistent with the maximum observed neutron star spin~\citep{Hessels:2006ze}, $\chi_{\max}=0.4$, while the second population restricts to $\chi_{\max}=0.05$, following the observed spins of Galactic double neutron stars~\citep{Lorimer:2008se, Lo:2010bj}. The mass distribution is chosen based on \cite{Farrow:2019xnc}; the mass ratio is drawn from a narrow truncated Gaussian with mean $\mu=1$, width $\sigma=0.1$, and lower limit $q_{\min}=0.4$. The total mass distribution is a power-law with index $\alpha=-2.5$ between $M_{\mathrm{tot,} \min}=2.3~M_{\odot}$ and $M_{\mathrm{tot,} \min}=4.3~M_{\odot}$ with low-mass smoothing (see Eq. 7-8 of \citealt{Talbot:2018cva}) over $\delta M_{\mathrm{tot}}=0.4~M_{\odot}$. We choose to parameterize the mass distribution in terms of total mass and mass ratio since the latter is particularly sensitive to correlations with the spin parameters~\citep{Purrer:2015nkh, Ng:2018neg}. For the extrinsic parameters, we assume the sources are distributed uniformly in comoving volume between luminosity distances of $d_{L}=10~\mathrm{Mpc}$ and $d_{L}=300~\mathrm{Mpc}$ and isotropically on the sky. Standard distributions are chosen for the remaining binary parameters (see e.g., \citealt{Romero-Shaw:2020owr}).

For both of the populations described above, we generate 100 events that are detectable with a LIGO Hanford-Livingston detector network~\citep{LIGOScientific:2014pky} operating at the sensitivity achieved during the third observing run~\citep{KAGRA:2013rdx, O3_psds}. We consider an event to be detectable if it is observed with a network optimal signal-to-noise ratio (SNR) $\rho_{\mathrm{opt}}^{\mathrm{net}}\geq 9$. For each event, we perform Bayesian parameter estimation to obtain samples from the posterior probability distributions for the binary parameters, $\boldsymbol{\theta}$, describing an individual BNS system, $i$:
\begin{align}
    p(\boldsymbol{\theta_{i}} | d_{i}) \propto \mathcal{L}(d_{i} | \boldsymbol{\theta_{i}}) \pi_{\mathrm{PE}}(\boldsymbol{\theta_{i}}),
    \label{eq:post}
\end{align}
where $\mathcal{L}(d_{i} | \boldsymbol{\theta_{i}})$ is the likelihood of observing data $d_{i}$ given the binary parameters $\boldsymbol{\theta_{i}}$~\citep{Veitch:2009hd, Romano:2016dpx}:
\begin{align}
    \mathcal{L}(d_{i} | \boldsymbol{\theta_{i}}) \propto \exp\left( -\sum_{k} \frac{2|d_{k}-h_{k}(\boldsymbol{\theta_{i}})|^{2}}{TS_{k}}\right).
    \label{eq:likelihood}
\end{align}
Here $h(\boldsymbol{\theta_{i}})$ represents the gravitational waveform for the BNS signal with parameters $\boldsymbol{\theta_{i}}$, $T$ is the duration of the analyzed data segment, $S_{k}$ is the noise power spectral density (PSD) characterizing the sensitivity of the detector, and $k$ indicates the frequency dependence of the data, waveform, and PSD. The prior in Eq.~\ref{eq:post} is denoted by $\pi_{\mathrm{PE}}(\boldsymbol{\theta_{i}})$, with the ``PE'' subscript indicating that this is the prior assumed during the initial parameter estimation step for each individual event.

We use the reduced order quadrature (ROQ) implementation~\citep{Smith:2016qas} of the IMRPhenomPv2 waveform model~\citep{Hannam:2013oca, Khan:2015jqa, Husa:2015iqa} in the likelihood above in order to curtail the computational cost of the individual-event parameter estimation. As such, we assume that the neutron stars we simulate are point masses with no tidal deformability. While previous works have demonstrated the importance of simultaneously inferring the mass and tidal deformability distributions at the risk of introducing biases when they are fit independently~\citep{Wysocki:2020myz, Golomb:2021tll}, we limit the scope of this work to mass-spin correlations and leave a full exploration of the potential correlations between the three intrinsic parameters of each neutron star to future studies. However, we comment on the effects of correlations with tides in Section~\ref{sec:conclusions}. We use the \textsc{PyMultiNest}~\citep{Feroz:2007kg, Feroz:2008xx, Feroz:2013hea, Buchner:2014nha} and \textsc{dynesty}~\citep{Speagle:2019ivv} nested samplers as implemented in the \textsc{Bilby}~\citep{Ashton:2018jfp, Romero-Shaw:2020owr} parameter estimation package to obtain samples from the posterior distributions for each event.

We choose priors $\pi_{\mathrm{PE}}(\boldsymbol{\theta_{i}})$ that are uniform in chirp mass with a width of $0.2~M_{\odot}$ centered on the true value of the chirp mass for each event and uniform in mass ratio over $[0.125, 1]$. This choice of prior and mass parameterization is more convenient for sampling. The priors on the extrinsic parameters are the same as those from which the populations were drawn. We adopt a number of different choices for the spin priors, as described below and summarized in Table~\ref{tab:runs}.

\begin{table}
	\centering
	\caption{Summary of the different combinations of true spin distribution and prior applied during the initial parameter estimation step for each of the scenarios we explore in this work. The $p_{\mathrm{draw}}$ column describes the distribution assumed for the sensitivity injections used to incorporate selection effects as detailed in Appendix~\ref{ap:injections}.}
	\label{tab:runs}
	\begin{tabular}{lll} %
		\hline
		$\pi_{\mathrm{pop}}(\boldsymbol{\chi_{1}}, \boldsymbol{\chi_{2}})$ & $\pi_{\mathrm{PE}}(\boldsymbol{\chi_{1}}, \boldsymbol{\chi_{2}})$ & $p_{\mathrm{draw}}(\boldsymbol{\chi_{1}}, \boldsymbol{\chi_{2}})$\\
		\hline
		Aligned, $\chi_{\max} = 0.4$ & Aligned, $\chi_{\max} = 0.8$ & Aligned, $\chi_{\max} = 0.8$ \\
		Aligned, $\chi_{\max} = 0.4$ & Aligned, $\chi_{\max} = 0.4$ & Aligned, $\chi_{\max} = 0.4$\\
		Aligned, $\chi_{\max} = 0.4$ & Aligned, $\chi_{\max} = 0.05$ & Aligned, $\chi_{\max} = 0.05$\\
		Aligned, $\chi_{\max} = 0.05$ & Aligned, $\chi_{\max} = 0.8$ & Aligned, $\chi_{\max} = 0.8$\\
		Aligned, $\chi_{\max} = 0.05$ & Precessing, $\chi_{\max} = 0.8$ & Aligned, $\chi_{\max} = 0.8$\\
		\hline
	\end{tabular}
\end{table}

Once we have obtained posterior samples for each of the individual BNS events, we can combine them to measure the underlying mass distribution of our simulated population. In this case, we are no longer interested in the individual binary parameters, $\boldsymbol{\theta_{i}}$, but rather in a set of hyperparameters $\boldsymbol{\Lambda}$ that describe the population distribution, $\pi_{\mathrm{pop}}(\boldsymbol{\theta}| \boldsymbol{\Lambda})$, which is also referred to as the hyper-prior. For our choice of population distribution, $\boldsymbol{\Lambda} = \{\alpha, M_{\mathrm{tot,} \min}, M_{\mathrm{tot,} \max}, \delta M_{\mathrm{tot}}, \mu, \sigma\}$. The likelihood of observing a set of individual events $\{d\}$ given the hyperparameters $\boldsymbol{\Lambda}$ is obtained by marginalizing over the binary parameters for each event and multiplying the resultant marginal likelihoods:
\begin{align}
    \mathcal{L}(\{d\} | \boldsymbol{\Lambda}) = \prod_{i}\int \mathcal{L}(d_{i} | \boldsymbol{\theta_{i}})\pi_{\mathrm{pop}}( \boldsymbol{\theta_{i}} | \boldsymbol{\Lambda})d\boldsymbol{\theta_{i}}.
\end{align}
This joint likelihood can be constructed from the individual-event posterior samples obtained in the first parameter estimation step via the ``recycling'' method (e.g. \citealt{Thrane:2018qnx}),
\begin{align}
    \mathcal{L}(\{d\} | \boldsymbol{\Lambda}) \propto \prod_{i}\sum_{j}\frac{\pi_{\mathrm{pop}}( \boldsymbol{\theta_{i,j}} | \boldsymbol{\Lambda})}{\pi_{\mathrm{PE}}(\boldsymbol{\theta_{i,j}})},
    \label{eq:hyper-like}
\end{align}
so that the joint likelihood is the ratio of the hyper-prior and the original PE prior, where the subscript $j$ denotes a sum over the individual posterior samples for each event. 

The likelihood above assumes that the individual events included in the observed population are an unbiased sample of the true population found in nature. However, we know that this is not the case for observed gravitational-wave events, since the detectors are more sensitive to high-mass sources~\citep{Fishbach:2017zga}. This selection effect needs to be accounted for in the likelihood in order to obtain unbiased estimates of the hyperparameters describing the astrophysical, rather than the observed, distribution~\citep{Loredo_2004, Mandel:2018mve, Thrane:2018qnx, Vitale:2020aaz},
\begin{align}
    \label{eq:vt_like}
    \mathcal{L}(\{d\} | \boldsymbol{\Lambda}) &= \prod_{i} \frac{\int \mathcal{L}(d_{i} | \boldsymbol{\theta_{i}})\pi_{\mathrm{pop}}(\boldsymbol{\Lambda} | \boldsymbol{\theta_{i}})d\boldsymbol{\theta_{i}}}{\alpha(\boldsymbol{\Lambda})},\\ 
    \alpha(\boldsymbol{\Lambda}) &= \int d\boldsymbol{\theta_{i}}p_{\mathrm{det}}(\boldsymbol{\theta_{i}})\pi_{\mathrm{pop}}(\boldsymbol{\theta_{i}} | \boldsymbol{\Lambda}).
    \label{eq:vt}
\end{align}
The function $p_{\mathrm{det}}(\boldsymbol{\theta_{i}})$ gives the probability that an individual event with parameters $\boldsymbol{\theta_{i}}$ will be detected. We evaluate $\alpha(\boldsymbol{\Lambda})$ using a Monte Carlo integral over a set of simulated signals drawn from the distribution $p_{\mathrm{draw}}(\boldsymbol{\theta})$ following the approach described in \citet{Farr:2019rap}. More details on the simulated population used for determining the detection probability can be found in Appendix~\ref{ap:injections}.
We evaluate the corrected likelihood in Eq.~\ref{eq:vt_like} to obtain samples from the posterior distributions of the hyperparameters using the \textsc{nestle} sampler~\citep{nestle} and the \textsc{GWPopulation} package~\citep{Talbot:2019okv}. The priors on the hyperparameters are all uniform over the ranges given in Table~\ref{tab:hyperparams}.

\begin{table}
	\centering
	\caption{Hyperparameters describing the mass distribution and the maximum and minimum values allowed in the prior applied during hierarchical inference. The priors on all parameters are uniform.}
	\label{tab:hyperparams}
	\begin{tabular}{lcll} %
		\hline
		Symbol & Parameter & Minimum & Maximum\\
		\hline
		$\alpha$ & total mass power-law index & 0 & 4\\
		$M_{\mathrm{tot,} \min}$ & minimum total mass & $2~M_{\odot}$ & $3~M_{\odot}$\\
		$M_{\mathrm{tot,} \max}$ & maximum total mass & $3.2~M_{\odot}$ & $5~M_{\odot}$\\
		$\delta M_{\mathrm{tot}}$ & smoothing parameter & $0~M_{\odot}$ & $1~M_{\odot}$\\
		$\mu$ & mass ratio mean & 0.4 & 1\\
		$\sigma$ & mass ratio standard deviation & 0.01 & 0.5\\
		\hline
	\end{tabular}
\end{table}

So far we have not mentioned the spins in the hierarchical inference step and have restricted the hyperparameters to include only those governing the mass distribution. For any parameters that are not explicitly included in the ratio in Eq.~\ref{eq:hyper-like}, it is implicitly assumed that the PE prior and the hyper-prior are the same (see Section 5.4 of \citealt{Vitale:2020aaz}). Even if we are not interested in measuring the hyperparameters of the spin distribution, if there is a mismatch between the underlying population distribution and the prior applied during the initial sampling, this can introduce biases in the hyperparameters that are being measured if there are significant correlations between those binary parameters, as is the case for the mass ratio and spins for binary neutron star systems, for example. To probe how such mismatches between the assumed and true population distributions for the spins can introduce biases in the mass distribution, we deliberately apply spin priors during the PE step that differ from those used to generate the population. This is consistent with what is currently done in population analyses, as will be described in detail in Section~\ref{sec:conclusions}.

\begin{figure}
	\includegraphics[width=\columnwidth]{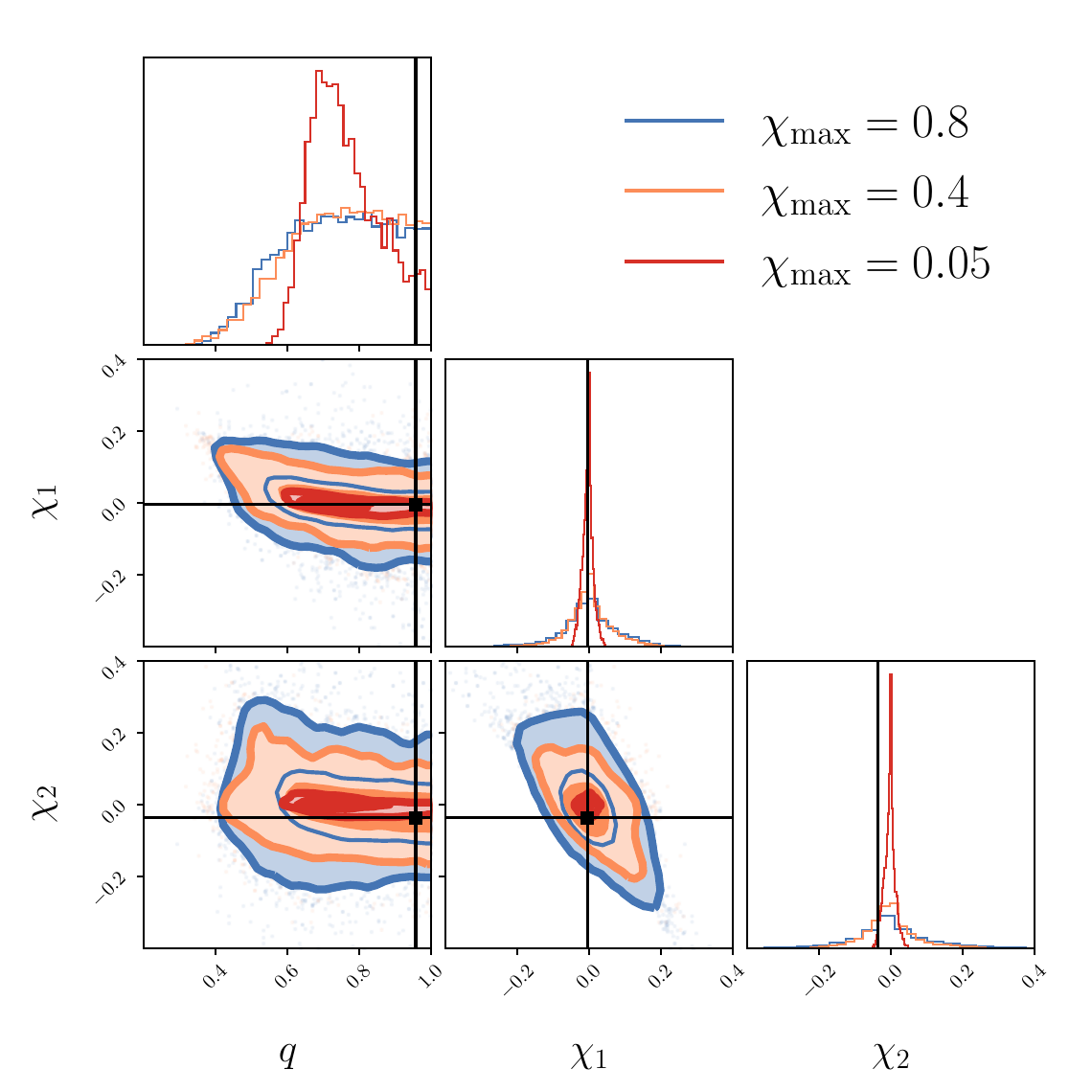}
    \caption{Corner plot showing the one-dimensional posteriors and 50 and 90\% contours for the mass ratio and component spin magnitudes for one event recovered with aligned-spin priors for three different choices of $\chi_{\max}$. The true values are indicated with the black lines and are $q=0.96, \chi_{1} = -0.005, \chi_{2} = -0.037$, within the prior range for all three choices of prior.}
    \label{fig:individual_corner}
\end{figure}

For the medium-spin population, we apply both high and low-spin priors with $\chi_{\max} = 0.8, 0.05$, respectively, as limited by the range of validity of the ROQ. We also perform parameter estimation and hierarchical inference on the mass distribution with a PE prior that matches the true population distribution with $\chi_{\max} = 0.4$ as a sanity check to ensure that the resultant biases we see are indeed due to a spin prior mismatch. An example corner plot showing the posteriors on the mass ratio and component spins obtained under each of these priors for one individual event is shown in Fig.~\ref{fig:individual_corner}. While the true values of all three parameters are included within the prior range for all three choices of spin prior, the $q$ posterior peaks at lower values when analyzed with the low-spin prior compared to the other two due to the correlation between mass ratio and spin.
For the low-spin population, we apply the high-spin prior first assuming aligned spins and then relax this assumption to allow for precessing spins. The combinations of true population and PE prior are summarized in Table~\ref{tab:runs}.

\section{Mass-spin correlations}
\label{sec:results}
\begin{figure}
	\includegraphics[width=\columnwidth]{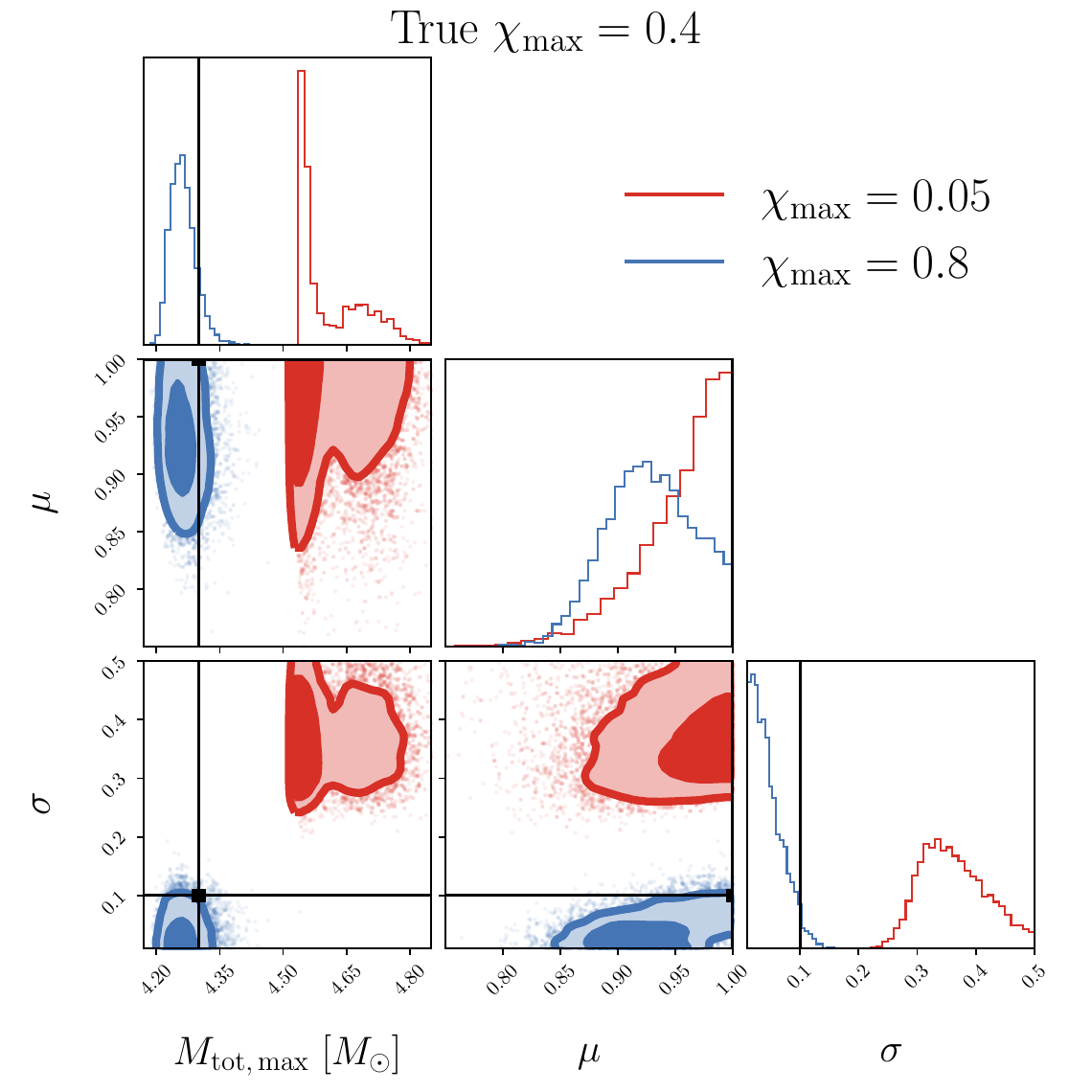}
    \caption{Hyperparameter posteriors on the maximum total mass and mean and width of the mass ratio distribution for two different choices of initial sampling prior applied to the medium-spin population. While the sources were drawn from an aligned-spin distribution with $\chi_{\max}=0.4$, the red and blue contours show the hierarchical inference results when they are recovered with priors with $\chi_{\max}=0.05, 0.8$, respectively. The black lines denote the true values of the hyperparameters. The value for $\mu=1$ lies at the edge of the prior, and hence the black line is not visible for this parameter. The high-spin prior results are conistent with the true population, while the low-spin prior results favor a much wider mass ratio distribution and a higher maximum total mass. The true values are excluded at $>3\sigma$ confidence.}
    \label{fig:chi_max_comp}
\end{figure}

The hyperparameter posteriors for $M_{\mathrm{tot}, \max}$, $\mu$, and $\sigma$ obtained when analyzing the medium-spin population with true $\chi_{\max}=0.4$ with both the low and high-spin priors are shown in the corner plot in Fig.~\ref{fig:chi_max_comp}. The blue contours show the results obtained with the high-spin prior. The true hyperparameter values are all recovered within the 90\% credible region of the posterior, demonstrating that analyzing a population with $\chi_{\max} = 0.4$ with a prior out to $\chi_{\max} = 0.8$ does not lead to biases in the inferred mass distribution. This is consistent with the fact that the analyses with the medium and high-spin priors shown in Fig.~\ref{fig:individual_corner} do not lead to significant differences in the mass ratio posteriors for individual events.

Conversely, the low-spin results shown in red are significantly biased in both $\sigma$ and $M_{\mathrm{tot}, \max}$. This can be explained by the correlations shown in Fig.~\ref{fig:individual_corner}, which are exacerbated by the low-spin prior and push the mass ratio posteriors for individual events towards lower values. This leads to a preference for wider distributions in the hierarchical inference step, since there is more support for extreme mass ratios in the population. Underestimating the mass ratio results in overestimating the total mass, since the chirp mass of the system,
\begin{align}
    \mathcal{M} = \left(\frac{q}{(1+q)^{2}}\right)^{3/5}M_{\mathrm{tot}},
\end{align}
is still well-constrained, propagating into the overestimation of the maximum total mass at the population level. 

\begin{figure*}
	\includegraphics[width=0.85\textwidth]{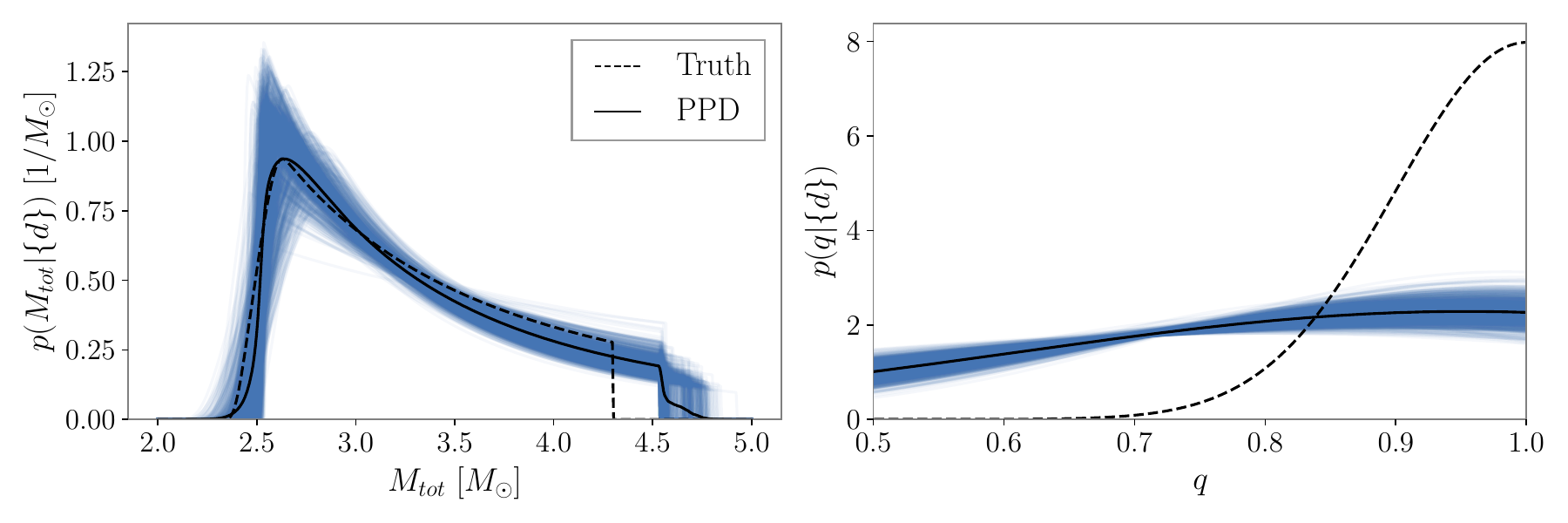}
	\includegraphics[width=0.85\textwidth]{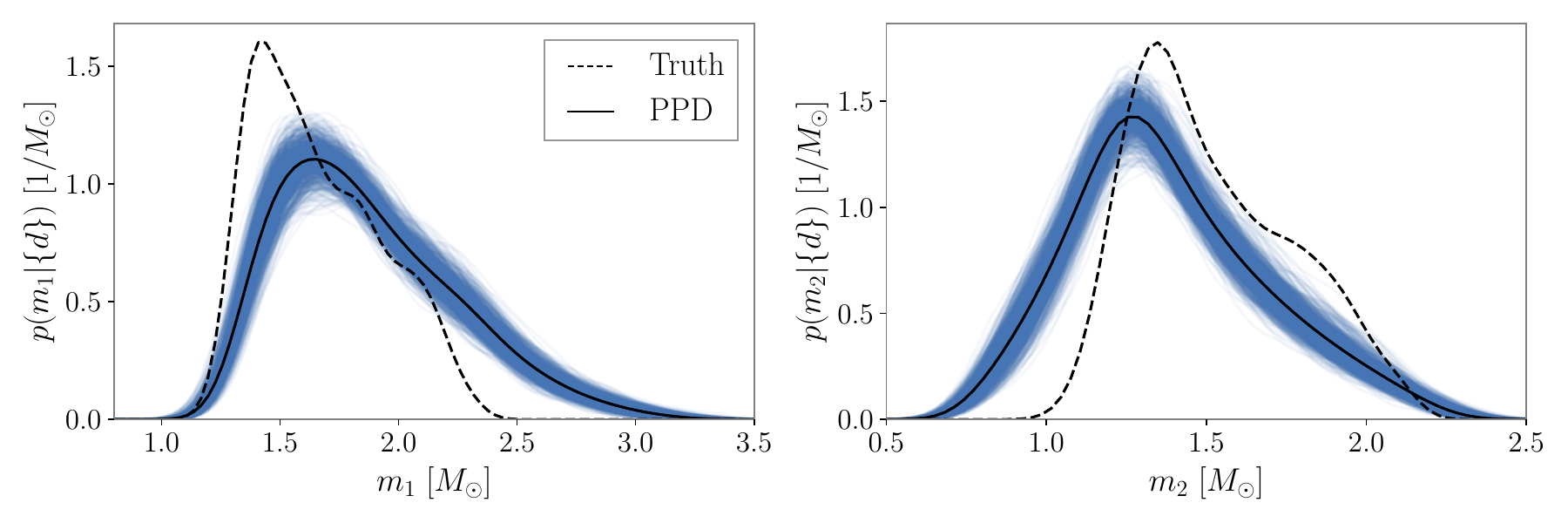}
    \caption{Inferred mass posterior population distributions (solid black lines) when a spin prior mismatch is introduced for a population where the true spin follows the aligned-spin distribution out to $\chi_{\max} = 0.4$, but the assumed population only allows $\chi_{\max} = 0.05$. The dashed black lines show the true distributions, while the light blue lines show individual draws from the hyperparameter posterior. Top: Total mass (left) and mass ratio (right) distributions. Bottom: Primary (left) and secondary (right) mass distributions.}
    \label{fig:chi04_low_spin_comp}
\end{figure*}

The inferred mass ratio and total mass distributions under the low-spin prior are shown in Fig.~\ref{fig:chi04_low_spin_comp}. These are represented by the posterior population distribution (PPD), which is the astrophysical distribution of the binary parameters $\boldsymbol{\theta}$ implied by the inferred hyperparameters $\boldsymbol{\Lambda}$~\citep{LIGOScientific:2020kqk}:
\begin{align}
    p(\boldsymbol{\theta} | \{d\}) = \int d\boldsymbol{\Lambda}p(\boldsymbol{\Lambda} | \{d\})\pi_{\mathrm{pop}}(\boldsymbol{\theta} | \boldsymbol{\Lambda}).
\end{align}
The recovery of a wider mass ratio distribution and higher maximum total mass results in overestimating the maximum mass of the primary neutron star and underestimating the minimum mass of the secondary, as can be seen in the bottom panel of Fig.~\ref{fig:chi04_low_spin_comp}. These biases can have profound implications for both single and binary neutron star formation mechanisms and their equation of state. Few equations of state are able to support neutron stars with $m\geq 2.5~M_{\odot}$, where 7\% of the probability lies for the recovered PPD on $m_{1}$. Additionally, it is difficult to form neutron stars with $m \lesssim 1~M_{\odot}$ under current stellar evolution models~\citep{Vigna-Gomez:2018dza}, a region of parameter space which contains 10\% of the probability for the inferred PPD on $m_{2}$. 

\begin{figure}
	\includegraphics[width=0.9\columnwidth]{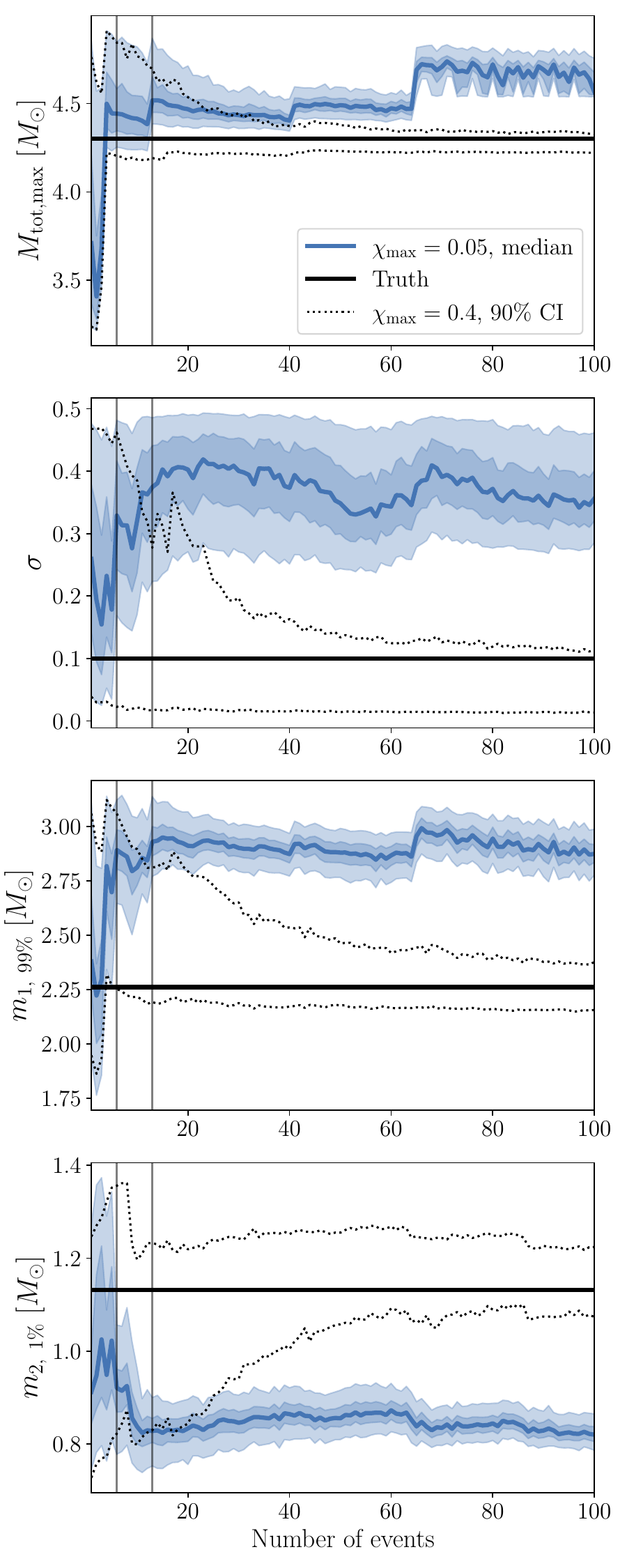}
    \caption{The top two panels show the evolution of the hyperparameters $M_{\mathrm{tot}, \max}$ and $\sigma$ as a function of the number of events included in the hierarchical analysis. The bottom two panels show the evolution of the maximum and minimum inferred component masses, represented by the 99th percentile of the $m_{1}$ distribution and the first percentile of the $m_{2}$ distribution, respectively. The solid black line shows the true value for each hyperparameter, while the solid blue line shows the median obtained when applying the low-spin prior to the medium-spin population. The blue shading gives the 50\% and 90\% credible intervals. The dotted black lines denote the 90\% credible region when there is no spin prior mismatch applied to the medium-spin population. The vertical grey lines show the sixth and thirteenth events, where the true values of $\sigma$ and $M_{\mathrm{tot, max}}$ are excluded at $>3\sigma$, respectively.
    }
    \label{fig:chi04_low_spin_constraints}
\end{figure}

The evolution of the bias in the hyperparameters as a function of the number of individual events included in the analysis is shown in blue in Fig.~\ref{fig:chi04_low_spin_constraints}. With six events, denoted by the first vertical grey line, the true value of $\sigma$ is already excluded from the $3\sigma$ credible region. The same occurs with 13 events for the maximum total mass parameter. We also show the evolution of the bias in the maximum and minimum components masses, which are represented by the 99th percentile of the primary mass distribution and the first percentile of the secondary mass distribution, respectively, since we do not directly parameterize the population in terms of the component masses. We find similar constraints on $m_{\max}$ and $m_{\min}$ to those presented in \citet{Chatziioannou:2020msi}, but these parameters become significantly biased with a similar number of events to the $\sigma$ and $M_{\mathrm{tot}, \max}$ hyperparameters. The bias in $M_{\mathrm{tot}, \max}$ is driven by four events in particular, which correspond to the distinguishable upward jumps in the top panel of Fig.~\ref{fig:chi04_low_spin_constraints}. These events all have true values of $|\boldsymbol{\chi_{1}}|>0.05$, which is outside the range allowed by the low-spin prior. This leads to a significant bias in the mass ratio posterior towards low values, driving the total mass upwards to keep the chirp mass constant. We emphasize that the posteriors for these individual events do not exhibit railing against the prior edges in either the spin or mass parameters and are well-converged. As such, they would not immediately be identified as problematic if they corresponded to real events. These results demonstrate that choosing a population model for the spins (even implicitly) that does not include all the sources in the population can significantly bias the BNS mass distribution. The width of the 90\% credible interval (CI) for the same parameters in the case of no spin-prior mismatch is shown in the black dotted lines in Fig.~\ref{fig:chi04_low_spin_constraints} for comparison. The true values of all the hyperparameters lie within the 90\% credible interval, demonstrating that there is no bias when the implied and true spin distributions match.

\begin{figure}
	\includegraphics[width=\columnwidth]{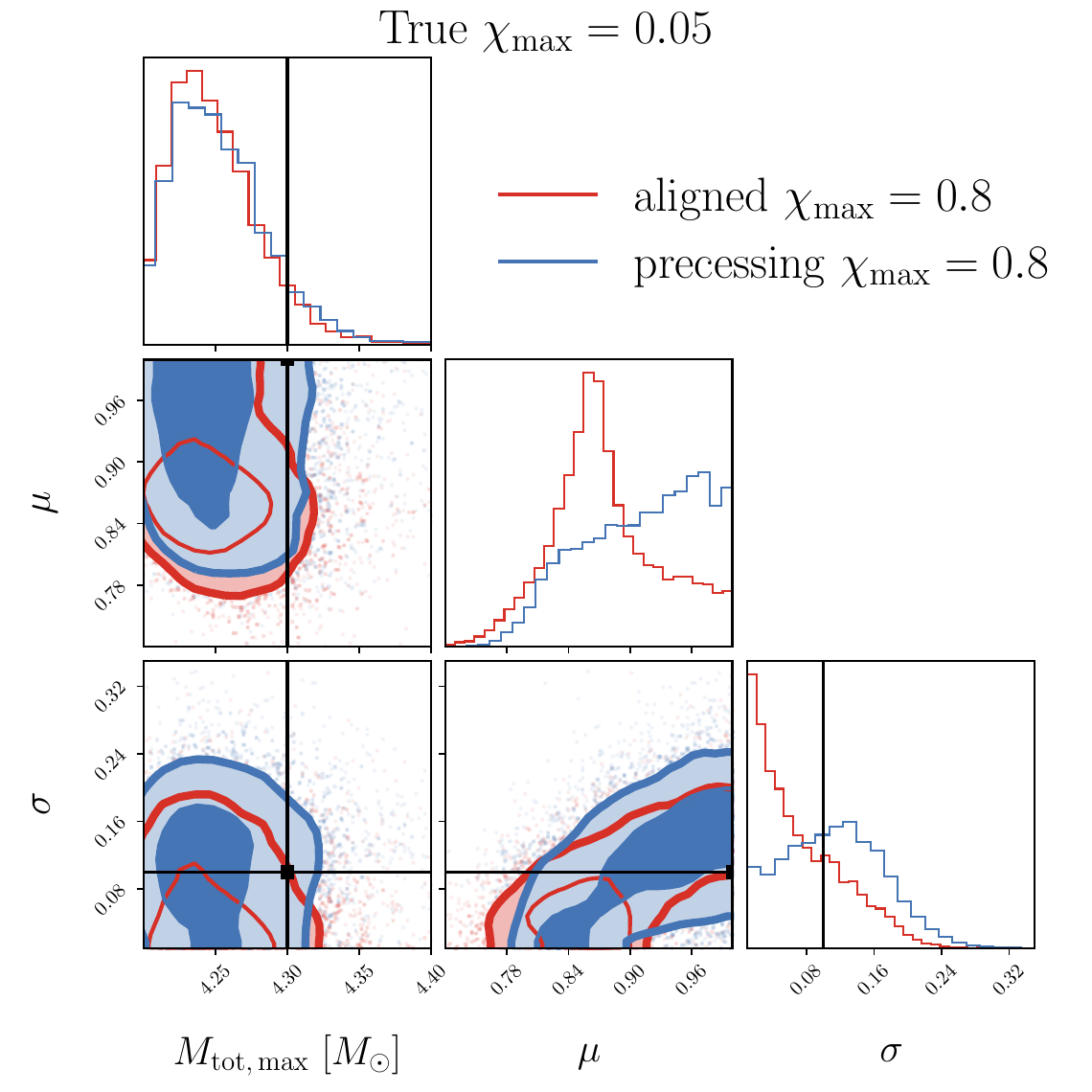}
    \caption{Hyperparameter posteriors on the maximum total mass and mean and width of the mass ratio distribution for two different choices of initial sampling prior applied to the low-spin population. The sources were drawn from an aligned-spin distribution with $\chi_{\max}=0.05$, but both priors assume $\chi_{\max} = 0.8$. However, the red contours show the hierarchical inference results when an aligned-spin prior is applied, while the blue allows for spin precession. While the high aligned-spin results are only marginally consistent with the true hyperparameter values, allowing for precession breaks the strong degeneracy between $q$ and $\chi_{z}$ and ameliorates the bias.}
    \label{fig:aligned_precessing_comp}
\end{figure}

While we have found that analyzing the population of sources with $\chi_{\max}=0.4$ with a prior going up to $\chi_{\max} = 0.8$ does not introduce a bias on the mass distribution, we now seek to investigate if the same conclusion holds for the low-spin population with $\chi_{\max} = 0.05$. The hyperparameter posteriors for the low-spin population analyzed with high-spin priors assuming both aligned and precessing spins are shown in the corner plot in Fig.~\ref{fig:aligned_precessing_comp}. The true values of the hyperparameters are not always contained within the 90\% credible region for the high aligned-spin prior shown in red, indicating a hint of a bias when 100 individual events are included in the population. The much larger difference in the prior volume between the low- and high-spin priors in this case leads to a bias even when all the observed events have spins within the allowed prior region. Allowing for the tilts to be misaligned to the orbital angular momentum introduces additional degrees of freedom that break the strong degeneracy between $q$ and $\chi_{z}$, alleviating this bias. The results obtained under the high precessing-spin prior shown in blue in Fig.~\ref{fig:aligned_precessing_comp} include the true hyperparameter values within the 90\% credible region. Thus, we conclude that using the high, precessing-spin prior is the safest choice for BNS systems if the mass and spin distributions are not modeled simultaneously during hierarchical inference.

\section{Conclusions}
\label{sec:conclusions}
In this work, we have demonstrated that introducing a mismatch between the true, underlying spin distribution for BNS systems observed in gravitational waves and that assumed when characterizing individual systems can lead to a bias in the inferred mass distribution at the population level. If the mass and spin distributions are not fit simultaneously, the implied population model for the spin distribution is the same as the prior used when conducting parameter estimation for individual sources. To investigate the effects of such a mismatch, we simulated two distinct populations of BNS sources, one with medium aligned spins out to $\chi_{\max}=0.4$ and the other with low aligned spins out to $\chi_{\max}=0.05$. 

The mass distribution inferred for the medium-spin population was significantly biased when the individual sources were analyzed with a low aligned-spin prior but unbiased when analyzed with a high aligned-spin prior. The bias is due to the degeneracy between the mass ratio and aligned spin components for BNS systems, which pushes the mass ratio posteriors for individual sources out towards more extreme values and also drives the total mass towards higher values. This translates into an overestimation of the maximum neutron star mass and and underestimation of the minimum mass, with adverse implications for both the inference of the nuclear equation of state and supernova mechanisms. The most massive neutron stars with posterior support in Fig.~\ref{fig:chi04_low_spin_comp} are only supported by the stiffest equations of state. The illusion of these high-mass neutron stars in the population would falsely populate the putative lower mass gap between the heaviest neutron stars and lightest black holes. Additionally, the false presence of a significant subpopulation of sub-solar mass compact objects would affect the inferred contribution of primordial sub-solar mass black holes to the dark matter density~\citep{LIGOScientific:2018glc, LIGOScientific:2019kan, Nitz:2021vqh, LIGOScientific:2021job}, as neutron stars with such low masses are not expected to form theoretically, and ground-based gravitational-wave detectors are not sensitive to the gravitational radiation from less massive compact objects such as white dwarfs. While in principle tidal effects could be used to distinguish sub-solar mass black holes from neutron stars, in practice gravitational-wave constraints on the tidal deformability are often weak (e.g. \citealt{LIGOScientific:2020aai, LIGOScientific:2021qlt}). 

The mass distribution inferred for the low-spin population demonstrated a hint of bias with the high aligned-spin prior, but this was alleviated when the individual sources were analyzed with a prior allowing for misaligned spin tilts. The extra degrees of freedom introduced by the precessing-spin model break the strong degeneracy between $q$ and $\chi_{z}$. These biases demonstrate the importance of fitting the mass and spin distributions simultaneously, to avoid implicitly mismodeling the spin distribution. However, if the mass distribution must be analyzed independently, we conclude that using a high, precessing-spin prior for the individual sources is the safest choice.

We note that the unbiased results we obtain in this demonstration are only robust if the choice of parameterized mass model used during the hierarchical inference step is physically realistic. If the assumed shape of the mass distribution does not match the underlying population, further biases can be introduced. However, this sort of mismatch is unlikely to affect the inferred maximum and minimum neutron star masses as significantly as the spin prior mismatch, so long as the parameterized population model covers the full range of allowed neutron star masses. This potential problem can be further ameliorated by fitting the mass distribution with several different hierarchical models, including for example the bimodal models favored by current observations and simulations (e.g.~\citealt{Antoniadis:2016hxz, Alsing:2017bbc, Tauris:2017omb, Vigna-Gomez:2018dza, Farrow:2019xnc}), and comparing the statistical evidence obtained between them to determine which provides the best fit. Alternatively, a more flexible model that does not impose a specific functional form on the mass distribution could be used~(e.g. \citealt{Mandel:2018mve, Wong:2020jdt,Tiwari:2020vym,Li:2021ukd, Sadiq:2021fin, Rinaldi:2021bhm}).

Correlations between the tidal parameters--which we have set to zero in our analysis--and the masses and spins can also introduce systematic errors in the inferred mass distribution if not accounted for. For the posteriors of individual events, more extreme values of mass ratio allow for smaller values of the tidal parameter which enters the gravitational waveform at leading order, $\tilde{\Lambda}$~\citep{Wade:2014vqa, LIGOScientific:2018hze}. Since the high-spin prior typically provides more support for more unequal mass ratios, changing the spin prior can also affect the inferred tidal parameters. Comparing our results with those of \cite{Golomb:2021tll}, which demonstrates the effect of mismodeling the tidal parameters on the inferred mass distribution, we conclude that enforcing a low-spin prior when there are larger spins in the population introduces a much more significant bias in the inferred mass distribution. However, both types of mismatches lead to increased support for higher neutron star masses.

In addition to the mismatches between the true and implied population models for the spins, masses, and tides discussed above, inadequate sampler performance when conducting parameter estimation for individual events can also manifest as a bias in the inferred population properties when multiple events are combined, as demonstrated in Appendix~\ref{ap:pp}. However, this sort of bias can be diagnosed and addressed by performing hierarchical inference on a simulated population.
Another potential source of systematic error is the accuracy limitations of the waveform models used to infer the properties of individual events. For the SNRs expected with the current generation of detectors, this effect should be small compared to the bias introduced by mismodeling the spin distribution~\citep{LIGOScientific:2018mvr, Dudi:2018jzn, Dietrich:2018uni, Messina:2019uby, LIGOScientific:2020aai}.

The upcoming fourth observing run of the LIGO and Virgo detectors is expected to add tens of new binary neutron star sources to the catalog of compact binaries detected in gravitational waves~\citep{KAGRA:2013rdx}. Based on our results, a bias in the mass distribution could be imposed on the observed population with as few as four additional BNS detections if a low-spin prior is applied to a population with higher spins. We note that while current population studies do not model the BNS mass distribution independently due to the paucity of detections, a variety of models consider BNS sources as part of the compact object population as a whole~\citep{Mandel:2016prl, Fishbach:2020ryj, Farah:2021qom, LIGOScientific:2021psn} or as part of the population of neutron star-containing systems~\citep{Landry:2021hvl, LIGOScientific:2021psn, 2021arXiv210806986L}. Most of these models and analyses assume the BNS spin distribution is uniform in magnitude on the interval [0, 1) with isotropic tilts, corresponding to the safe choice identified in Section~\ref{sec:results}. 
Both the \textsc{Multi Source} model presented in \citet{LIGOScientific:2021psn} and the analysis of \citet{2021arXiv210806986L} fit the mass and spin distributions of the BNS subpopulation simultaneously. The \textsc{Multi Source} model assumes the spin magnitudes follow a Beta distribution with $\chi_{\max}=0.05$ and the tilts are isotropically distributed. The analysis of \citet{2021arXiv210806986L} takes a similar approach, assuming the neutron star spin magnitudes follow a truncated Gaussian out to $\chi_{\max}=0.05$ and fitting the tilts as a mixture model between aligned and isotropic distributions. While assuming small spins will not lead to a bias for the two BNS events currently detected, this assumption should be relaxed to avoid introducing a bias in the mass distribution with the first few events detected during the next observing run.

\section*{Acknowledgements}
S.B., C.T. and S.V.\ acknowledge support of the National Science Foundation and the LIGO Laboratory.
LIGO was constructed by the California Institute of Technology and
Massachusetts Institute of Technology with funding from the National
Science Foundation and operates under cooperative agreement PHY-0757058.
S.B. is also supported by the NSF Graduate Research Fellowship under Grant No. DGE-1122374.
S.V. is also supported by NSF PHY-2045740.
The authors would like to thank Amanda Farah and Phil Landry for insightful comments on the manuscript.
The authors are grateful for computational resources provided by the LIGO Lab and supported by NSF Grants PHY-0757058 and PHY-0823459.
This paper carries LIGO document number LIGO-P2100426.

\section*{Data Availability}

A public repository with example code to reproduce the results in this manuscript is available on \href{https://github.com/asb5468/bns_mass_distribution}{github}. Only simulated data were analyzed using the publicly-available programs \textsc{Bilby}~\citep{Ashton:2018jfp, Romero-Shaw:2020owr}, \textsc{PyMultiNest}~\citep{Feroz:2007kg, Feroz:2008xx, Feroz:2013hea, Buchner:2014nha}, \textsc{dynesty}~\citep{Speagle:2019ivv}, \textsc{nestle}~\citep{nestle}, and \textsc{GWPopulation}~\citep{Talbot:2019okv}

\bibliographystyle{mnras}
\bibliography{mnras} %

\begin{thebibliography}{}
\makeatletter
\relax
\def\mn@urlcharsother{\let\do\@makeother \do\$\do\&\do\#\do\^\do\_\do\%\do\~}
\def\mn@doi{\begingroup\mn@urlcharsother \@ifnextchar [ {\mn@doi@}
  {\mn@doi@[]}}
\def\mn@doi@[#1]#2{\def\@tempa{#1}\ifx\@tempa\@empty \href
  {http://dx.doi.org/#2} {doi:#2}\else \href {http://dx.doi.org/#2} {#1}\fi
  \endgroup}
\def\mn@eprint#1#2{\mn@eprint@#1:#2::\@nil}
\def\mn@eprint@arXiv#1{\href {http://arxiv.org/abs/#1} {{\tt arXiv:#1}}}
\def\mn@eprint@dblp#1{\href {http://dblp.uni-trier.de/rec/bibtex/#1.xml}
  {dblp:#1}}
\def\mn@eprint@#1:#2:#3:#4\@nil{\def\@tempa {#1}\def\@tempb {#2}\def\@tempc
  {#3}\ifx \@tempc \@empty \let \@tempc \@tempb \let \@tempb \@tempa \fi \ifx
  \@tempb \@empty \def\@tempb {arXiv}\fi \@ifundefined
  {mn@eprint@\@tempb}{\@tempb:\@tempc}{\expandafter \expandafter \csname
  mn@eprint@\@tempb\endcsname \expandafter{\@tempc}}}

\bibitem[\protect\citeauthoryear{Aasi et~al.}{Aasi
  et~al.}{2015}]{LIGOScientific:2014pky}
Aasi J.,  et~al., 2015, \mn@doi [Class. Quant. Grav.]
  {10.1088/0264-9381/32/7/074001}, 32, 074001

\bibitem[\protect\citeauthoryear{Abbott et~al.}{Abbott
  et~al.}{2017}]{LIGOScientific:2017vwq}
Abbott B.~P.,  et~al., 2017, \mn@doi [Phys. Rev. Lett.]
  {10.1103/PhysRevLett.119.161101}, 119, 161101

\bibitem[\protect\citeauthoryear{Abbott et~al.}{Abbott
  et~al.}{2018a}]{KAGRA:2013rdx}
Abbott B.~P.,  et~al., 2018a, \mn@doi [Living Rev. Rel.]
  {10.1007/s41114-020-00026-9}, 21, 3

\bibitem[\protect\citeauthoryear{Abbott et~al.}{Abbott
  et~al.}{2018b}]{LIGOScientific:2017zlf}
Abbott B.~P.,  et~al., 2018b, \mn@doi [Phys. Rev. Lett.]
  {10.1103/PhysRevLett.120.091101}, 120, 091101

\bibitem[\protect\citeauthoryear{Abbott et~al.}{Abbott
  et~al.}{2018c}]{LIGOScientific:2018glc}
Abbott B.~P.,  et~al., 2018c, \mn@doi [Phys. Rev. Lett.]
  {10.1103/PhysRevLett.121.231103}, 121, 231103

\bibitem[\protect\citeauthoryear{Abbott et~al.}{Abbott
  et~al.}{2019a}]{LIGOScientific:2018hze}
Abbott B.~P.,  et~al., 2019a, \mn@doi [Phys. Rev. X]
  {10.1103/PhysRevX.9.011001}, 9, 011001

\bibitem[\protect\citeauthoryear{Abbott et~al.}{Abbott
  et~al.}{2019b}]{LIGOScientific:2018mvr}
Abbott B.~P.,  et~al., 2019b, \mn@doi [Phys. Rev. X]
  {10.1103/PhysRevX.9.031040}, 9, 031040

\bibitem[\protect\citeauthoryear{Abbott et~al.}{Abbott
  et~al.}{2019c}]{LIGOScientific:2019kan}
Abbott B.~P.,  et~al., 2019c, \mn@doi [Phys. Rev. Lett.]
  {10.1103/PhysRevLett.123.161102}, 123, 161102

\bibitem[\protect\citeauthoryear{Abbott et~al.}{Abbott et~al.}{2020a}]{O3_psds}
Abbott B.~P.,  et~al., 2020a, Noise curves used for Simulations in the update
  of the Observing Scenarios Paper, https://dcc.ligo.org/LIGO-T2000012/public,
  \url {https://dcc.ligo.org/LIGO-T2000012/public}

\bibitem[\protect\citeauthoryear{Abbott et~al.}{Abbott
  et~al.}{2020b}]{LIGOScientific:2020aai}
Abbott B.~P.,  et~al., 2020b, \mn@doi [Astrophys. J. Lett.]
  {10.3847/2041-8213/ab75f5}, 892, L3

\bibitem[\protect\citeauthoryear{Abbott et~al.}{Abbott
  et~al.}{2021a}]{LIGOScientific:2021job}
Abbott R.,  et~al., 2021a, arXiv e-prints, \href
  {https://ui.adsabs.harvard.edu/abs/2021arXiv210912197T} {p. arXiv:2109.12197}

\bibitem[\protect\citeauthoryear{Abbott et~al.}{Abbott
  et~al.}{2021b}]{LIGOScientific:2021djp}
Abbott R.,  et~al., 2021b, arXiv e-prints, \href
  {https://ui.adsabs.harvard.edu/abs/2021arXiv211103606T} {p. arXiv:2111.03606}

\bibitem[\protect\citeauthoryear{Abbott et~al.}{Abbott
  et~al.}{2021c}]{LIGOScientific:2021psn}
Abbott R.,  et~al., 2021c, arXiv e-prints, \href
  {https://ui.adsabs.harvard.edu/abs/2021arXiv211103634T} {p. arXiv:2111.03634}

\bibitem[\protect\citeauthoryear{Abbott et~al.}{Abbott
  et~al.}{2021d}]{LIGOScientific:2020kqk}
Abbott R.,  et~al., 2021d, \mn@doi [Astrophys. J. Lett.]
  {10.3847/2041-8213/abe949}, 913, L7

\bibitem[\protect\citeauthoryear{Abbott et~al.}{Abbott
  et~al.}{2021e}]{LIGOScientific:2021qlt}
Abbott R.,  et~al., 2021e, \mn@doi [Astrophys. J. Lett.]
  {10.3847/2041-8213/ac082e}, 915, L5

\bibitem[\protect\citeauthoryear{Alsing \& Handley}{Alsing \&
  Handley}{2021}]{Alsing:2021wef}
Alsing J.,  Handley W.,  2021, \mn@doi [Mon. Not. Roy. Astron. Soc.]
  {10.1093/mnrasl/slab057}, 505, L95

\bibitem[\protect\citeauthoryear{Alsing, Silva  \& Berti}{Alsing
  et~al.}{2018}]{Alsing:2017bbc}
Alsing J.,  Silva H.~O.,   Berti E.,  2018, \mn@doi [Mon. Not. Roy. Astron.
  Soc.] {10.1093/mnras/sty1065}, 478, 1377

\bibitem[\protect\citeauthoryear{{Antoniadis}, {Tauris}, {Ozel}, {Barr},
  {Champion}  \& {Freire}}{{Antoniadis} et~al.}{2016}]{Antoniadis:2016hxz}
{Antoniadis} J.,  {Tauris} T.~M.,  {Ozel} F.,  {Barr} E.,  {Champion} D.~J.,
  {Freire} P. C.~C.,  2016, arXiv e-prints, \href
  {https://ui.adsabs.harvard.edu/abs/2016arXiv160501665A} {p. arXiv:1605.01665}

\bibitem[\protect\citeauthoryear{Ashton et~al.}{Ashton
  et~al.}{2019}]{Ashton:2018jfp}
Ashton G.,  et~al., 2019, \mn@doi [Astrophys. J. Suppl.]
  {10.3847/1538-4365/ab06fc}, 241, 27

\bibitem[\protect\citeauthoryear{Barbary et~al.}{Barbary et~al.}{2021}]{nestle}
Barbary K.,  et~al., 2021, Nestle, \url{https://github.com/kbarbary/nestle}

\bibitem[\protect\citeauthoryear{Berry et~al.}{Berry
  et~al.}{2015}]{Berry:2014jja}
Berry C. P.~L.,  et~al., 2015, \mn@doi [Astrophys. J.]
  {10.1088/0004-637X/804/2/114}, 804, 114

\bibitem[\protect\citeauthoryear{Bouffanais, Mapelli, Santoliquido, Giacobbo,
  Di~Carlo, Rastello, Artale  \& Iorio}{Bouffanais
  et~al.}{2021}]{Bouffanais:2021wcr}
Bouffanais Y.,  Mapelli M.,  Santoliquido F.,  Giacobbo N.,  Di~Carlo U.~N.,
  Rastello S.,  Artale M.~C.,   Iorio G.,  2021, \mn@doi [\mnras]
  {10.1093/mnras/stab2438}, \href
  {https://ui.adsabs.harvard.edu/abs/2021MNRAS.507.5224B} {507, 5224}

\bibitem[\protect\citeauthoryear{Buchner et~al.,}{Buchner
  et~al.}{2014}]{Buchner:2014nha}
Buchner J.,  et~al., 2014, \mn@doi [Astron. Astrophys.]
  {10.1051/0004-6361/201322971}, 564, A125

\bibitem[\protect\citeauthoryear{Campanelli, Lousto  \& Zlochower}{Campanelli
  et~al.}{2006}]{Campanelli:2006uy}
Campanelli M.,  Lousto C.~O.,   Zlochower Y.,  2006, \mn@doi [Phys. Rev. D]
  {10.1103/PhysRevD.74.041501}, 74, 041501

\bibitem[\protect\citeauthoryear{Chatziioannou \& Farr}{Chatziioannou \&
  Farr}{2020}]{Chatziioannou:2020msi}
Chatziioannou K.,  Farr W.~M.,  2020, \mn@doi [Phys. Rev. D]
  {10.1103/PhysRevD.102.064063}, 102, 064063

\bibitem[\protect\citeauthoryear{Chernoff \& Finn}{Chernoff \&
  Finn}{1993}]{Chernoff:1993th}
Chernoff D.~F.,  Finn L.~S.,  1993, \mn@doi [Astrophys. J. Lett.]
  {10.1086/186898}, 411, L5

\bibitem[\protect\citeauthoryear{Cook, Gelman  \& Rubin}{Cook
  et~al.}{2006}]{Cook:2006}
Cook S.~R.,  Gelman A.,   Rubin D.~B.,  2006, \mn@doi [Journal of Computational
  and Graphical Statistics] {10.1198/106186006X136976}, 15, 675

\bibitem[\protect\citeauthoryear{Cutler \& Flanagan}{Cutler \&
  Flanagan}{1994}]{Cutler:1994ys}
Cutler C.,  Flanagan E.~E.,  1994, \mn@doi [Phys. Rev. D]
  {10.1103/PhysRevD.49.2658}, 49, 2658

\bibitem[\protect\citeauthoryear{Dietrich et~al.}{Dietrich
  et~al.}{2019}]{Dietrich:2018uni}
Dietrich T.,  et~al., 2019, \mn@doi [Phys. Rev. D]
  {10.1103/PhysRevD.99.024029}, 99, 024029

\bibitem[\protect\citeauthoryear{Dudi, Pannarale, Dietrich, Hannam, Bernuzzi,
  Ohme  \& Br\"ugmann}{Dudi et~al.}{2018}]{Dudi:2018jzn}
Dudi R.,  Pannarale F.,  Dietrich T.,  Hannam M.,  Bernuzzi S.,  Ohme F.,
  Br\"ugmann B.,  2018, \mn@doi [Phys. Rev. D] {10.1103/PhysRevD.98.084061},
  98, 084061

\bibitem[\protect\citeauthoryear{{Farah}, {Fishbach}, {Essick}, {Holz}  \&
  {Galaudage}}{{Farah} et~al.}{2021}]{Farah:2021qom}
{Farah} A.~M.,  {Fishbach} M.,  {Essick} R.,  {Holz} D.~E.,   {Galaudage} S.,
  2021, arXiv e-prints, \href
  {https://ui.adsabs.harvard.edu/abs/2021arXiv211103498F} {p. arXiv:2111.03498}

\bibitem[\protect\citeauthoryear{Farr}{Farr}{2019}]{Farr:2019rap}
Farr W.~M.,  2019, \mn@doi [Research Notes of the AAS]
  {10.3847/2515-5172/ab1d5f}, 3, 66

\bibitem[\protect\citeauthoryear{Farr et~al.}{Farr et~al.}{2016}]{Farr:2015lna}
Farr B.,  et~al., 2016, \mn@doi [Astrophys. J.] {10.3847/0004-637X/825/2/116},
  825, 116

\bibitem[\protect\citeauthoryear{Farrow, Zhu  \& Thrane}{Farrow
  et~al.}{2019}]{Farrow:2019xnc}
Farrow N.,  Zhu X.-J.,   Thrane E.,  2019, \mn@doi [Astrophys. J.]
  {10.3847/1538-4357/ab12e3}, 876, 18

\bibitem[\protect\citeauthoryear{Feroz \& Hobson}{Feroz \&
  Hobson}{2008}]{Feroz:2007kg}
Feroz F.,  Hobson M.~P.,  2008, \mn@doi [Mon. Not. Roy. Astron. Soc.]
  {10.1111/j.1365-2966.2007.12353.x}, 384, 449

\bibitem[\protect\citeauthoryear{Feroz, Hobson  \& Bridges}{Feroz
  et~al.}{2009}]{Feroz:2008xx}
Feroz F.,  Hobson M.~P.,   Bridges M.,  2009, \mn@doi [Mon. Not. Roy. Astron.
  Soc.] {10.1111/j.1365-2966.2009.14548.x}, 398, 1601

\bibitem[\protect\citeauthoryear{Feroz, Hobson, Cameron  \& Pettitt}{Feroz
  et~al.}{2019}]{Feroz:2013hea}
Feroz F.,  Hobson M.~P.,  Cameron E.,   Pettitt A.~N.,  2019, \mn@doi [Open J.
  Astrophys.] {10.21105/astro.1306.2144}, 2, 10

\bibitem[\protect\citeauthoryear{{Finke}, {Foffa}, {Iacovelli}, {Maggiore}  \&
  {Mancarella}}{{Finke} et~al.}{2021}]{Finke:2021eio}
{Finke} A.,  {Foffa} S.,  {Iacovelli} F.,  {Maggiore} M.,   {Mancarella} M.,
  2021, arXiv e-prints, \href
  {https://ui.adsabs.harvard.edu/abs/2021arXiv210804065F} {p. arXiv:2108.04065}

\bibitem[\protect\citeauthoryear{Finn}{Finn}{1996}]{Finn:1995ah}
Finn L.~S.,  1996, \mn@doi [Phys. Rev. D] {10.1103/PhysRevD.53.2878}, 53, 2878

\bibitem[\protect\citeauthoryear{Fishbach \& Holz}{Fishbach \&
  Holz}{2017}]{Fishbach:2017zga}
Fishbach M.,  Holz D.~E.,  2017, \mn@doi [Astrophys. J. Lett.]
  {10.3847/2041-8213/aa9bf6}, 851, L25

\bibitem[\protect\citeauthoryear{Fishbach, Essick  \& Holz}{Fishbach
  et~al.}{2020}]{Fishbach:2020ryj}
Fishbach M.,  Essick R.,   Holz D.~E.,  2020, \mn@doi [Astrophys. J. Lett.]
  {10.3847/2041-8213/aba7b6}, 899, L8

\bibitem[\protect\citeauthoryear{Galaudage, Adamcewicz, Zhu, Stevenson  \&
  Thrane}{Galaudage et~al.}{2021}]{Galaudage:2020zst}
Galaudage S.,  Adamcewicz C.,  Zhu X.-J.,  Stevenson S.,   Thrane E.,  2021,
  \mn@doi [Astrophys. J. Lett.] {10.3847/2041-8213/abe7f6}, 909, L19

\bibitem[\protect\citeauthoryear{{Golomb} \& {Talbot}}{{Golomb} \&
  {Talbot}}{2021}]{Golomb:2021tll}
{Golomb} J.,  {Talbot} C.,  2021, arXiv e-prints, \href
  {https://ui.adsabs.harvard.edu/abs/2021arXiv210615745G} {p. arXiv:2106.15745}

\bibitem[\protect\citeauthoryear{Hannam, Brown, Fairhurst, Fryer  \&
  Harry}{Hannam et~al.}{2013}]{Hannam:2013uu}
Hannam M.,  Brown D.~A.,  Fairhurst S.,  Fryer C.~L.,   Harry I.~W.,  2013,
  \mn@doi [Astrophys. J. Lett.] {10.1088/2041-8205/766/1/L14}, 766, L14

\bibitem[\protect\citeauthoryear{Hannam, Schmidt, Boh\'e, Haegel, Husa, Ohme,
  Pratten  \& P\"urrer}{Hannam et~al.}{2014}]{Hannam:2013oca}
Hannam M.,  Schmidt P.,  Boh\'e A.,  Haegel L.,  Husa S.,  Ohme F.,  Pratten
  G.,   P\"urrer M.,  2014, \mn@doi [Phys. Rev. Lett.]
  {10.1103/PhysRevLett.113.151101}, 113, 151101

\bibitem[\protect\citeauthoryear{Hessels, Ransom, Stairs, Freire, Kaspi  \&
  Camilo}{Hessels et~al.}{2006}]{Hessels:2006ze}
Hessels J. W.~T.,  Ransom S.~M.,  Stairs I.~H.,  Freire P. C.~C.,  Kaspi V.~M.,
    Camilo F.,  2006, \mn@doi [Science] {10.1126/science.1123430}, 311, 1901

\bibitem[\protect\citeauthoryear{Husa, Khan, Hannam, P\"urrer, Ohme,
  Jim\'enez~Forteza  \& Boh\'e}{Husa et~al.}{2016}]{Husa:2015iqa}
Husa S.,  Khan S.,  Hannam M.,  P\"urrer M.,  Ohme F.,  Jim\'enez~Forteza X.,
  Boh\'e A.,  2016, \mn@doi [Phys. Rev. D] {10.1103/PhysRevD.93.044006}, 93,
  044006

\bibitem[\protect\citeauthoryear{Khan, Husa, Hannam, Ohme, P\"urrer,
  Jim\'enez~Forteza  \& Boh\'e}{Khan et~al.}{2016}]{Khan:2015jqa}
Khan S.,  Husa S.,  Hannam M.,  Ohme F.,  P\"urrer M.,  Jim\'enez~Forteza X.,
  Boh\'e A.,  2016, \mn@doi [Phys. Rev. D] {10.1103/PhysRevD.93.044007}, 93,
  044007

\bibitem[\protect\citeauthoryear{Kiziltan, Kottas, De~Yoreo  \&
  Thorsett}{Kiziltan et~al.}{2013}]{Kiziltan:2013oja}
Kiziltan B.,  Kottas A.,  De~Yoreo M.,   Thorsett S.~E.,  2013, \mn@doi
  [Astrophys. J.] {10.1088/0004-637X/778/1/66}, 778, 66

\bibitem[\protect\citeauthoryear{{Landry} \& {Read}}{{Landry} \&
  {Read}}{2021}]{Landry:2021hvl}
{Landry} P.,  {Read} J.~S.,  2021, arXiv e-prints, \href
  {https://ui.adsabs.harvard.edu/abs/2021arXiv210704559L} {p. arXiv:2107.04559}

\bibitem[\protect\citeauthoryear{Landry, Essick  \& Chatziioannou}{Landry
  et~al.}{2020}]{Landry:2020vaw}
Landry P.,  Essick R.,   Chatziioannou K.,  2020, \mn@doi [Phys. Rev. D]
  {10.1103/PhysRevD.101.123007}, 101, 123007

\bibitem[\protect\citeauthoryear{Lasota, Haensel  \& Abramowicz}{Lasota
  et~al.}{1996}]{Lasota:1995eu}
Lasota J.-P.,  Haensel P.,   Abramowicz M.~A.,  1996, \mn@doi [Astrophys. J.]
  {10.1086/176650}, 456, 300

\bibitem[\protect\citeauthoryear{Legred, Chatziioannou, Essick, Han  \&
  Landry}{Legred et~al.}{2021}]{Legred:2021hdx}
Legred I.,  Chatziioannou K.,  Essick R.,  Han S.,   Landry P.,  2021, \mn@doi
  [Phys. Rev. D] {10.1103/PhysRevD.104.063003}, 104, 063003

\bibitem[\protect\citeauthoryear{{Li}, {Tang}, {Wang}, {Yuan}, {Fan}  \&
  {Wei}}{{Li} et~al.}{2021a}]{2021arXiv210806986L}
{Li} Y.-J.,  {Tang} S.-P.,  {Wang} Y.-Z.,  {Yuan} Q.,  {Fan} Y.-Z.,   {Wei}
  D.-M.,  2021a, arXiv e-prints, \href
  {https://ui.adsabs.harvard.edu/abs/2021arXiv210806986L} {p. arXiv:2108.06986}

\bibitem[\protect\citeauthoryear{Li, Wang, Han, Tang, Yuan, Fan  \& Wei}{Li
  et~al.}{2021b}]{Li:2021ukd}
Li Y.-J.,  Wang Y.-Z.,  Han M.-Z.,  Tang S.-P.,  Yuan Q.,  Fan Y.-Z.,   Wei
  D.-M.,  2021b, \mn@doi [Astrophys. J.] {10.3847/1538-4357/ac0971}, 917, 33

\bibitem[\protect\citeauthoryear{Lo \& Lin}{Lo \& Lin}{2011}]{Lo:2010bj}
Lo K.-W.,  Lin L.-M.,  2011, \mn@doi [Astrophys. J.]
  {10.1088/0004-637X/728/1/12}, 728, 12

\bibitem[\protect\citeauthoryear{Loredo}{Loredo}{2004}]{Loredo_2004}
Loredo T.~J.,  2004, in Bayesian Inference and Maximum Entropy Methods in
  Science and Engineering ed R. Fischer, R. Preuss and U. V. Toussaint. AIP,
  Melville, NY, p.~195, \mn@doi{10.1063/1.1835214}

\bibitem[\protect\citeauthoryear{Lorimer}{Lorimer}{2008}]{Lorimer:2008se}
Lorimer D.~R.,  2008, \mn@doi [Living Rev. Rel.] {10.12942/lrr-2008-8}, 11, 8

\bibitem[\protect\citeauthoryear{Mandel}{Mandel}{2010}]{Mandel:2009pc}
Mandel I.,  2010, \mn@doi [Phys. Rev. D] {10.1103/PhysRevD.81.084029}, 81,
  084029

\bibitem[\protect\citeauthoryear{Mandel, Farr, Colonna, Stevenson, Ti\v{n}o  \&
  Veitch}{Mandel et~al.}{2017}]{Mandel:2016prl}
Mandel I.,  Farr W.~M.,  Colonna A.,  Stevenson S.,  Ti\v{n}o P.,   Veitch J.,
  2017, \mn@doi [Mon. Not. Roy. Astron. Soc.] {10.1093/mnras/stw2883}, 465,
  3254

\bibitem[\protect\citeauthoryear{Mandel, Farr  \& Gair}{Mandel
  et~al.}{2019}]{Mandel:2018mve}
Mandel I.,  Farr W.~M.,   Gair J.~R.,  2019, \mn@doi [Mon. Not. Roy. Astron.
  Soc.] {10.1093/mnras/stz896}, 486, 1086

\bibitem[\protect\citeauthoryear{Messina, Dudi, Nagar  \& Bernuzzi}{Messina
  et~al.}{2019}]{Messina:2019uby}
Messina F.,  Dudi R.,  Nagar A.,   Bernuzzi S.,  2019, \mn@doi [Phys. Rev. D]
  {10.1103/PhysRevD.99.124051}, 99, 124051

\bibitem[\protect\citeauthoryear{Miller, Chirenti  \& Lamb}{Miller
  et~al.}{2019}]{Miller:2019nzo}
Miller M.~C.,  Chirenti C.,   Lamb F.~K.,  2019, \mn@doi [Astrophys. J.]
  {10.3847/1538-4357/ab4ef9}, 888, 12

\bibitem[\protect\citeauthoryear{Mukherjee, Parkinson  \& Liddle}{Mukherjee
  et~al.}{2006}]{Mukherjee:2005wg}
Mukherjee P.,  Parkinson D.,   Liddle A.~R.,  2006, \mn@doi [Astrophys. J.
  Lett.] {10.1086/501068}, 638, L51

\bibitem[\protect\citeauthoryear{Ng, Vitale, Zimmerman, Chatziioannou, Gerosa
  \& Haster}{Ng et~al.}{2018}]{Ng:2018neg}
Ng K. K.~Y.,  Vitale S.,  Zimmerman A.,  Chatziioannou K.,  Gerosa D.,   Haster
  C.-J.,  2018, \mn@doi [Phys. Rev. D] {10.1103/PhysRevD.98.083007}, 98, 083007

\bibitem[\protect\citeauthoryear{Nitz \& Wang}{Nitz \&
  Wang}{2021}]{Nitz:2021vqh}
Nitz A.~H.,  Wang Y.-F.,  2021, \mn@doi [Phys. Rev. Lett.]
  {10.1103/PhysRevLett.127.151101}, 127, 151101

\bibitem[\protect\citeauthoryear{\"Ozel \& Freire}{\"Ozel \&
  Freire}{2016}]{Ozel:2016oaf}
\"Ozel F.,  Freire P.,  2016, \mn@doi [Ann. Rev. Astron. Astrophys.]
  {10.1146/annurev-astro-081915-023322}, 54, 401

\bibitem[\protect\citeauthoryear{Ozel, Psaltis, Narayan  \& Villarreal}{Ozel
  et~al.}{2012}]{Ozel:2012ax}
Ozel F.,  Psaltis D.,  Narayan R.,   Villarreal A.~S.,  2012, \mn@doi
  [Astrophys. J.] {10.1088/0004-637X/757/1/55}, 757, 55

\bibitem[\protect\citeauthoryear{Pankow}{Pankow}{2018}]{Pankow:2018iab}
Pankow C.,  2018, \mn@doi [Astrophys. J.] {10.3847/1538-4357/aadc66}, 866, 60

\bibitem[\protect\citeauthoryear{Pejcha, Thompson  \& Kochanek}{Pejcha
  et~al.}{2012}]{Pejcha:2012at}
Pejcha O.,  Thompson T.~A.,   Kochanek C.~S.,  2012, \mn@doi [Mon. Not. Roy.
  Astron. Soc.] {10.1111/j.1365-2966.2012.21369.x}, 424, 1570

\bibitem[\protect\citeauthoryear{P\"urrer, Hannam  \& Ohme}{P\"urrer
  et~al.}{2016}]{Purrer:2015nkh}
P\"urrer M.,  Hannam M.,   Ohme F.,  2016, \mn@doi [Phys. Rev. D]
  {10.1103/PhysRevD.93.084042}, 93, 084042

\bibitem[\protect\citeauthoryear{Rinaldi \& Del~Pozzo}{Rinaldi \&
  Del~Pozzo}{2021}]{Rinaldi:2021bhm}
Rinaldi S.,  Del~Pozzo W.,  2021, \mn@doi [Mon. Not. Roy. Astron. Soc.]
  {10.1093/mnras/stab3224}, 509, 5454

\bibitem[\protect\citeauthoryear{Romano \& Cornish}{Romano \&
  Cornish}{2017}]{Romano:2016dpx}
Romano J.~D.,  Cornish N.~J.,  2017, \mn@doi [Living Rev. Rel.]
  {10.1007/s41114-017-0004-1}, 20, 2

\bibitem[\protect\citeauthoryear{Romero-Shaw et~al.}{Romero-Shaw
  et~al.}{2020}]{Romero-Shaw:2020owr}
Romero-Shaw I.~M.,  et~al., 2020, \mn@doi [Mon. Not. Roy. Astron. Soc.]
  {10.1093/mnras/staa2850}, 499, 3295

\bibitem[\protect\citeauthoryear{Roulet, Chia, Olsen, Dai, Venumadhav, Zackay
  \& Zaldarriaga}{Roulet et~al.}{2021}]{Roulet:2021hcu}
Roulet J.,  Chia H.~S.,  Olsen S.,  Dai L.,  Venumadhav T.,  Zackay B.,
  Zaldarriaga M.,  2021, \mn@doi [Phys. Rev. D] {10.1103/PhysRevD.104.083010},
  104, 083010

\bibitem[\protect\citeauthoryear{{Sadiq}, {Dent}  \& {Wysocki}}{{Sadiq}
  et~al.}{2021}]{Sadiq:2021fin}
{Sadiq} J.,  {Dent} T.,   {Wysocki} D.,  2021, arXiv e-prints, \href
  {https://ui.adsabs.harvard.edu/abs/2021arXiv211212659S} {p. arXiv:2112.12659}

\bibitem[\protect\citeauthoryear{Safarzadeh, Ramirez-Ruiz  \&
  Berger}{Safarzadeh et~al.}{2020}]{Safarzadeh:2020efa}
Safarzadeh M.,  Ramirez-Ruiz E.,   Berger E.,  2020, \mn@doi [Astrophys. J.]
  {10.3847/1538-4357/aba596}, 900, 13

\bibitem[\protect\citeauthoryear{Shao, Tang, Jiang  \& Fan}{Shao
  et~al.}{2020}]{Shao:2020bzt}
Shao D.-S.,  Tang S.-P.,  Jiang J.-L.,   Fan Y.-Z.,  2020, \mn@doi [Phys. Rev.
  D] {10.1103/PhysRevD.102.063006}, 102, 063006

\bibitem[\protect\citeauthoryear{Shaw, Bridges  \& Hobson}{Shaw
  et~al.}{2007}]{Shaw:2007jj}
Shaw R.,  Bridges M.,   Hobson M.~P.,  2007, \mn@doi [Mon. Not. Roy. Astron.
  Soc.] {10.1111/j.1365-2966.2007.11871.x}, 378, 1365

\bibitem[\protect\citeauthoryear{Smith, Field, Blackburn, Haster, P\"urrer,
  Raymond  \& Schmidt}{Smith et~al.}{2016}]{Smith:2016qas}
Smith R.,  Field S.~E.,  Blackburn K.,  Haster C.-J.,  P\"urrer M.,  Raymond
  V.,   Schmidt P.,  2016, \mn@doi [Phys. Rev. D] {10.1103/PhysRevD.94.044031},
  94, 044031

\bibitem[\protect\citeauthoryear{Speagle}{Speagle}{2020}]{Speagle:2019ivv}
Speagle J.~S.,  2020, \mn@doi [Mon. Not. Roy. Astron. Soc.]
  {10.1093/mnras/staa278}, 493, 3132

\bibitem[\protect\citeauthoryear{Talbot \& Thrane}{Talbot \&
  Thrane}{2018}]{Talbot:2018cva}
Talbot C.,  Thrane E.,  2018, \mn@doi [Astrophys. J.]
  {10.3847/1538-4357/aab34c}, 856, 173

\bibitem[\protect\citeauthoryear{Talbot, Smith, Thrane  \& Poole}{Talbot
  et~al.}{2019}]{Talbot:2019okv}
Talbot C.,  Smith R.,  Thrane E.,   Poole G.~B.,  2019, \mn@doi [Phys. Rev. D]
  {10.1103/PhysRevD.100.043030}, 100, 043030

\bibitem[\protect\citeauthoryear{{Talts}, {Betancourt}, {Simpson}, {Vehtari}
  \& {Gelman}}{{Talts} et~al.}{2018}]{Talts:2018}
{Talts} S.,  {Betancourt} M.,  {Simpson} D.,  {Vehtari} A.,   {Gelman} A.,
  2018, arXiv e-prints, \href
  {https://ui.adsabs.harvard.edu/abs/2018arXiv180406788T} {p. arXiv:1804.06788}

\bibitem[\protect\citeauthoryear{Tauris et~al.}{Tauris
  et~al.}{2017}]{Tauris:2017omb}
Tauris T.~M.,  et~al., 2017, \mn@doi [Astrophys. J.]
  {10.3847/1538-4357/aa7e89}, 846, 170

\bibitem[\protect\citeauthoryear{Taylor \& Gair}{Taylor \&
  Gair}{2012}]{Taylor:2012db}
Taylor S.~R.,  Gair J.~R.,  2012, \mn@doi [Phys. Rev. D]
  {10.1103/PhysRevD.86.023502}, 86, 023502

\bibitem[\protect\citeauthoryear{Taylor, Gair  \& Mandel}{Taylor
  et~al.}{2012}]{Taylor:2011fs}
Taylor S.~R.,  Gair J.~R.,   Mandel I.,  2012, \mn@doi [Phys. Rev. D]
  {10.1103/PhysRevD.85.023535}, 85, 023535

\bibitem[\protect\citeauthoryear{Thorsett \& Chakrabarty}{Thorsett \&
  Chakrabarty}{1999}]{Thorsett:1998uc}
Thorsett S.~E.,  Chakrabarty D.,  1999, \mn@doi [Astrophys. J.]
  {10.1086/306742}, 512, 288

\bibitem[\protect\citeauthoryear{Thrane \& Talbot}{Thrane \&
  Talbot}{2019}]{Thrane:2018qnx}
Thrane E.,  Talbot C.,  2019, \mn@doi [Publ. Astron. Soc. Austral.]
  {10.1017/pasa.2019.2}, 36, e010

\bibitem[\protect\citeauthoryear{Tiwari}{Tiwari}{2021}]{Tiwari:2020vym}
Tiwari V.,  2021, \mn@doi [Class. Quant. Grav.] {10.1088/1361-6382/ac0b54}, 38,
  155007

\bibitem[\protect\citeauthoryear{Veitch \& Vecchio}{Veitch \&
  Vecchio}{2010}]{Veitch:2009hd}
Veitch J.,  Vecchio A.,  2010, \mn@doi [Phys. Rev. D]
  {10.1103/PhysRevD.81.062003}, 81, 062003

\bibitem[\protect\citeauthoryear{Vigna-G\'omez et~al.}{Vigna-G\'omez
  et~al.}{2018}]{Vigna-Gomez:2018dza}
Vigna-G\'omez A.,  et~al., 2018, \mn@doi [Mon. Not. Roy. Astron. Soc.]
  {10.1093/mnras/sty2463}, 481, 4009

\bibitem[\protect\citeauthoryear{{Vitale}, {Gerosa}, {Farr}  \&
  {Taylor}}{{Vitale} et~al.}{2020}]{Vitale:2020aaz}
{Vitale} S.,  {Gerosa} D.,  {Farr} W.~M.,   {Taylor} S.~R.,  2020, arXiv
  e-prints, \href {https://ui.adsabs.harvard.edu/abs/2020arXiv200705579V} {p.
  arXiv:2007.05579}

\bibitem[\protect\citeauthoryear{Wade, Creighton, Ochsner, Lackey, Farr,
  Littenberg  \& Raymond}{Wade et~al.}{2014}]{Wade:2014vqa}
Wade L.,  Creighton J. D.~E.,  Ochsner E.,  Lackey B.~D.,  Farr B.~F.,
  Littenberg T.~B.,   Raymond V.,  2014, \mn@doi [Phys. Rev. D]
  {10.1103/PhysRevD.89.103012}, 89, 103012

\bibitem[\protect\citeauthoryear{Wong, Contardo  \& Ho}{Wong
  et~al.}{2020}]{Wong:2020jdt}
Wong K. W.~K.,  Contardo G.,   Ho S.,  2020, \mn@doi [Phys. Rev. D]
  {10.1103/PhysRevD.101.123005}, 101, 123005

\bibitem[\protect\citeauthoryear{Wong, Breivik, Kremer  \& Callister}{Wong
  et~al.}{2021}]{Wong:2020ise}
Wong K. W.~K.,  Breivik K.,  Kremer K.,   Callister T.,  2021, \mn@doi [Phys.
  Rev. D] {10.1103/PhysRevD.103.083021}, 103, 083021

\bibitem[\protect\citeauthoryear{{Wysocki}, {O'Shaughnessy}, {Wade}  \&
  {Lange}}{{Wysocki} et~al.}{2020}]{Wysocki:2020myz}
{Wysocki} D.,  {O'Shaughnessy} R.,  {Wade} L.,   {Lange} J.,  2020, arXiv
  e-prints, \href {https://ui.adsabs.harvard.edu/abs/2020arXiv200101747W} {p.
  arXiv:2001.01747}

\bibitem[\protect\citeauthoryear{Zevin et~al.,}{Zevin
  et~al.}{2021}]{Zevin:2020gbd}
Zevin M.,  et~al., 2021, \mn@doi [Astrophys. J.] {10.3847/1538-4357/abe40e},
  910, 152

\bibitem[\protect\citeauthoryear{Zhu, Howell, Blair  \& Zhu}{Zhu
  et~al.}{2013}]{Zhu:2012xw}
Zhu X.-J.,  Howell E.~J.,  Blair D.~G.,   Zhu Z.-H.,  2013, \mn@doi [Mon. Not.
  Roy. Astron. Soc.] {10.1093/mnras/stt207}, 431, 882

\bibitem[\protect\citeauthoryear{{Zhu}, {Thrane}, {Os{\l}owski}, {Levin}  \&
  {Lasky}}{{Zhu} et~al.}{2018}]{2018PhRvD..98d3002Z}
{Zhu} X.,  {Thrane} E.,  {Os{\l}owski} S.,  {Levin} Y.,   {Lasky} P.~D.,  2018,
  \mn@doi [\prd] {10.1103/PhysRevD.98.043002}, \href
  {https://ui.adsabs.harvard.edu/abs/2018PhRvD..98d3002Z} {98, 043002}

\makeatother
\end{thebibliography}

\appendix
\section{The Insufficiency of P-P plots as a diagnostic test}
\label{ap:pp}
In order to obtain unbiased posteriors for the hyperparameters describing a population of events using the likelihood in Eq.~\ref{eq:vt_like}, the individual-event posteriors must also be unbiased. A common diagnostic tool for evaluating the performance of a stochastic sampler is a ``probability-probability plot``, or a P-P plot~\citep{Cook:2006, Talts:2018}. This provides a graphical way to verify that the true parameter values are recovered within a certain credible interval for the expected fraction of events in a population. If the likelihood correctly describes the distribution of data, the true parameter values should be recovered within the 5\% credible interval 5\% of the time, the 95\% credible interval 95\% of the time, etc. For the Whittle likelihood in Eq.~\ref{eq:likelihood} used when performing parameter estimation on individual gravitational-wave sources, this is satisfied if the data are Gaussian about the assumed noise power spectral density. In the case of unbiased individual-event posterior samples for a particular parameter, the P-P plot should be approximately diagonal, as it shows the fraction of events for which the true value of a given parameter falls within the given credible interval as a function of that credible interval. 

\begin{figure}
	\includegraphics[width=\columnwidth]{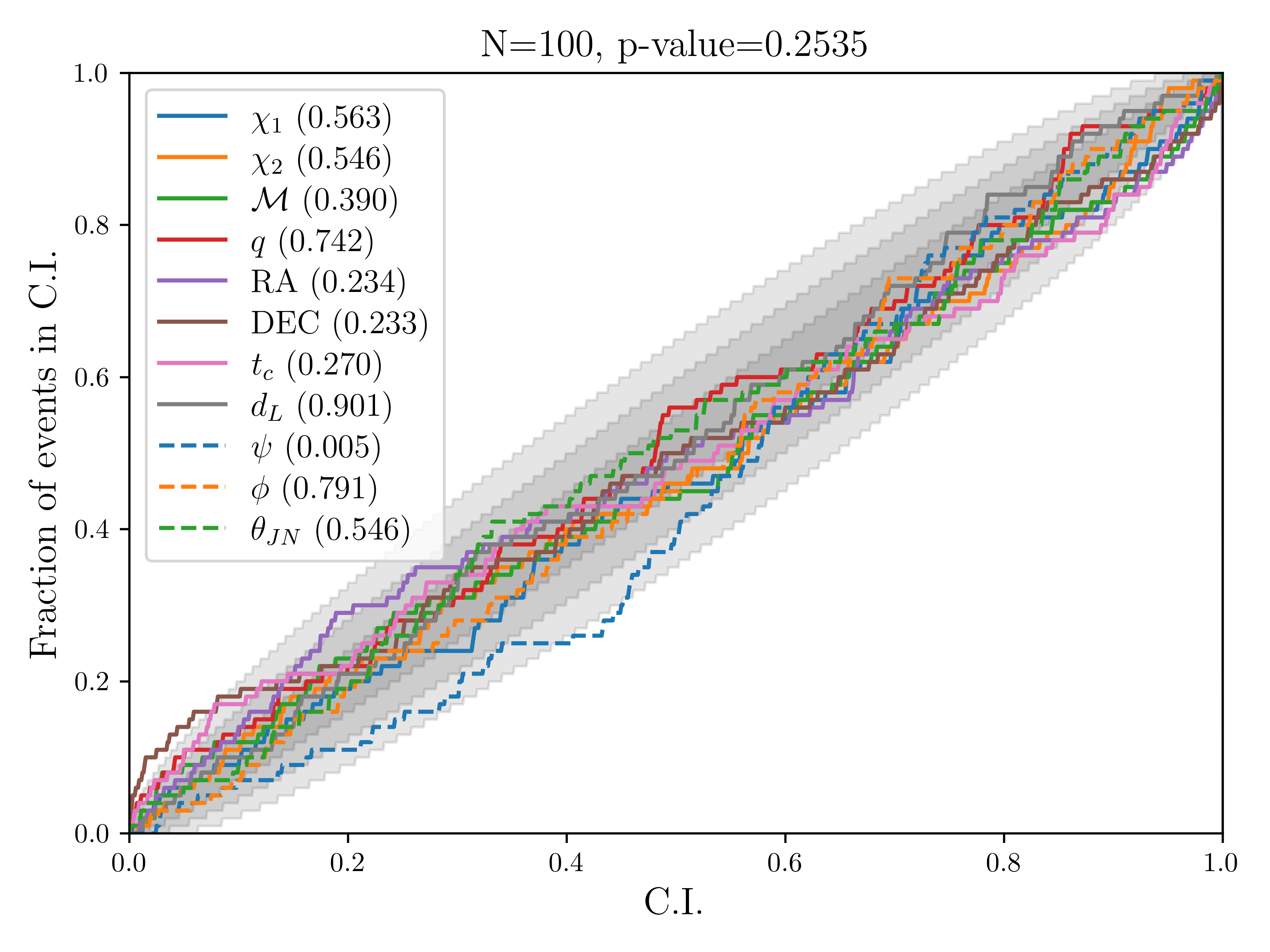}
    \caption{P-P plot showing the fraction of events for which the true value is recovered within a certain credible interval as a function of that credible interval for a population of 100 simulated BNS sources, sampled with the \textsc{PyMultiNest} package. The lines for individual parameters stay within the $3\sigma$ credible region, shaded in light gray, and the probability values quoted in the legend are consistent with passing the P-P test. The other grey shaded areas show the 1 and $2\sigma$ credible regions.}
    \label{fig:pp}
\end{figure}

An example P-P plot obtained using the \textsc{PyMultiNest} sampler for a population of 100 simulated BNS sources is shown in Fig.~\ref{fig:pp}. The extrinsic parameters are drawn from the same prior distributions described in the main text, and the spins follow the low aligned-spin prior. The mass ratio is also drawn from the same population prior explored in the main text, namely a narrow truncated Gaussian with $\mu =1,\ \sigma=0.1$. The chirp masses are drawn from a uniform prior between 1.52 and $1.70~M_{\odot}$. For the P-P plot to be unbiased, the distributions from which the events are drawn must match the priors applied during sampling, meaning that there is no prior mismatch and no cut based on the SNR of the individual events, as has been the case in the rest of this work. The legend shows the probability for the fraction of events within each credible interval to be drawn from a uniform distribution for individual parameters, as expected from Gaussianity. For the mass ratio, this value is $p=0.742$, passing the P-P test (where the threshold is $p>1/11=0.09$). The probability that the individual-parameter probabilities are drawn from a uniform distribution is 0.254, consistent with random chance for 11 parameters and indicative of unbiased sampling across the 11 parameters drawn from and recovered with this particular set of priors.

\begin{figure}
	\includegraphics[width=\columnwidth]{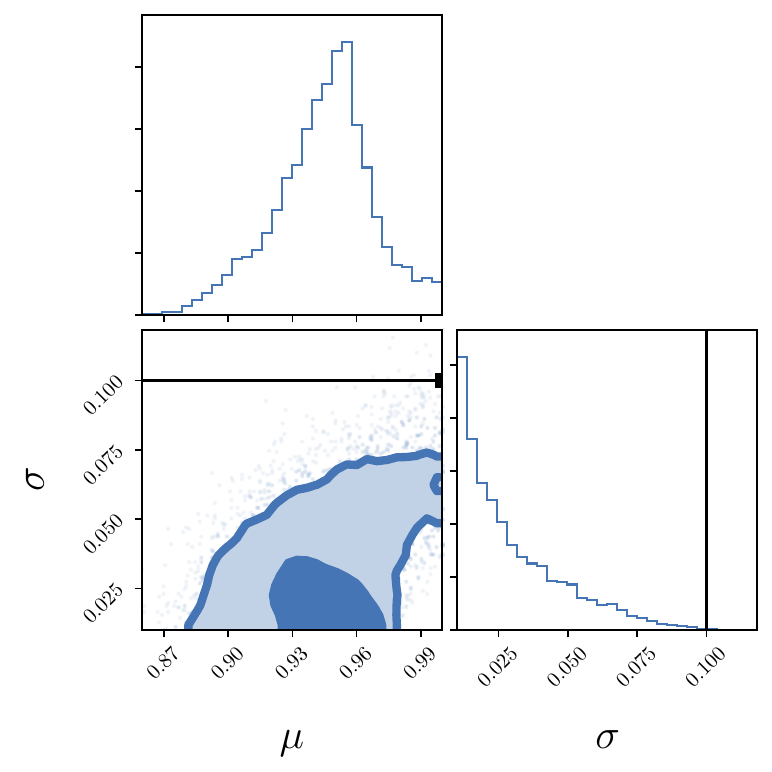}
    \caption{Corner plot for the inferred mass ratio hyperparameters using the 100 events used to generate the P-P plot in Fig.~\ref{fig:pp}.}
    \label{fig:infer_q}
\end{figure}

However, when the same 100 individual events are analyzed hierarchically to recover the true values of $\mu$ and $\sigma$, the results are biased at the $3\sigma$ level, as shown in Fig.~\ref{fig:infer_q}. This demonstrates a case where the sampling algorithm passes the P-P test for a particular population but still yields biased hierarchical inference results, highlighting the insufficiency of P-P plots as a diagnostic tool for individual-event parameter estimation. In this case, the recovered hyperparameters favor a narrowly distributed population peaking away from $\mu=1$, indicating that the sampler is unable to thoroughly explore the edge of the prior space where most of the probability lies for the nearly equal-mass events included in the population. This is due to the adapted simultaneous ellipsoidal nested sampling method used by \textsc{PyMultiNest}~\citep{Mukherjee:2005wg, Shaw:2007jj, Feroz:2007kg, Feroz:2008xx}, which bounds the iso-likelihood contours around clusters of live points with N-dimensional ellipsoids. Because the probability for mass ratio rails against the edge of the prior and the algorithm is inefficient at sampling near edges, the peak of the distribution at equal mass is undersampled.

\begin{figure}
	\includegraphics[width=\columnwidth]{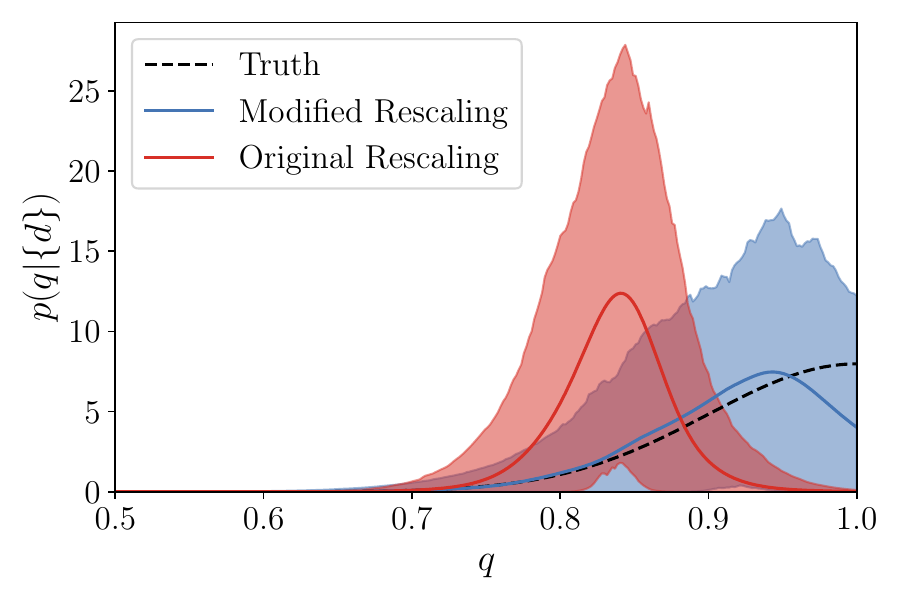}
    \caption{Inferred mass ratio PPDs when applying the original inverse CDF rescaling method (red) and the modified rescaling method described in Eq.~\ref{eq:rescaling} (blue) to the medium-spin population described in the main text analyzed with no spin prior mismatch. The dotted line shows the true distribution, and the shading shows the 90\% credible region. Without the modified rescaling, the mass ratio distribution is biased towards lower $q$ values and more narrowly peaked.}
    \label{fig:reflected_ppd}
\end{figure}

\textsc{PyMultiNest} internally works with a uniform prior for samples from the unit hyper-cube which must then be scaled to the physical parameter space such that the scaled samples are drawn from the desired physical prior distribution.
In order to improve the convergence near equal masses, we propose to use a two-stage mapping that shifts the peak of the probability at $q=1$ away from the edge of the prior in the frame of the sampler. Typically samples from the unit cube are rescaled onto the appropriate prior distribution for a given parameter via the inverse of the prior's cumulative distribution function, such that for a sample from the unit cube, $x$, $q(x) = \mathrm{CDF}^{-1}(x)$ (although other methods have been proposed, e.g.,~\citealt{Alsing:2021wef}). Here, we propose to add an intermediate step,
\begin{align}
    u &= 2\min(x, 1-x),\\
    q &= \mathrm{CDF}^{-1}(u), 
    \label{eq:rescaling}
\end{align}
where $u$ still takes on values within the unit interval, but instead of $q(x=1) = 1$, $q(x=0.5) = 1$. This transformation maps equal mass--where the peak of the probability lies--to the center of the sampled space rather than the edge, which is more difficult to sample. In Fig.~\ref{fig:reflected_ppd}, we show the mass ratio PPDs for the medium-spin population described in the main text with no spin prior mismatch obtained with and without this modified rescaling method. Similarly to the corner plot in Fig.~\ref{fig:infer_q}, without the modified rescaling, the PPD shown in red peaks at lower mass ratios and is more narrowly distributed. The true hyperparameters are excluded from the recovered posteriors at $>3\sigma$ credibility. Once the modified rescaling is implemented for the individual-event parameter estimation, the hierarchical inference becomes unbiased as shown in blue. This stealth bias introduced by the stochastic sampling algorithm that is not caught with a P-P test demonstrates the importance of verifying hierarchical inference analyses with synthetic populations where the true hyperparameter values are known and controllable before conducting the analysis on real data.

\section{Selection Effects}
\label{ap:injections}
In order to evaluate the selection function in Eq.~\ref{eq:vt}, we calculate the detection probability, $p_{\mathrm{det}}(\boldsymbol{\theta_{i}})$, using an injection campaign. $\alpha(\boldsymbol{\Lambda})$ gives the fraction of signals  that will be detected drawn from a population model with hyperparameters $\boldsymbol{\Lambda}$. We generate 194953 total simulated signals, calculating the network optimal SNR, $\rho_{\mathrm{opt}}^{\mathrm{net}}$ for each. Of these, 40000 of are above the threshold for detection, $\rho_{\mathrm{opt}}^{\mathrm{net}}\geq 9$. The distribution from which the injections are drawn, $p_{\mathrm{draw}}(\boldsymbol{\theta})$, is uniform in total mass over the range $[2,5]~M_{\odot}$, and uses the same prior distributions described in the main text for all the extrinsic parameters. The mass ratio distribution is
\begin{align}
    p_{\mathrm{draw}}(q) = 0.5\bigg( 1.67 &+ \frac{1}{\sigma}\frac{\mathcal{N}(q | \mu, \sigma)}{\Phi(\frac{q_{\max} - \mu}{\sigma}) - \Phi(\frac{q_{\min} - \mu}{\sigma})}\bigg), \\
    &q_{\min} < q < q_{\max} \nonumber
\end{align}
which is the normalized superposition of a uniform distribution and truncated Gaussian distribution with $\mu=1, \sigma=0.1$ between $q_{\min}=0.4$ and $q_{\max}=1$. The truncated Gaussian is added to enhance the number of injections with nearly equal mass ratios, since this is the part of the parameter space that should have the most support given the true distribution we used for the simulated populations described in Sec.~\ref{sec:methods}. The spins are drawn following the aligned-spin prior with $\chi_{\max}=0.99$. The injections are reweighted so that $\chi_{\max}$ matches the corresponding value used during the first parameter estimation step for each of the spin-prior mismatches we consider above. The different combinations of true population distribution, PE prior, and injection distribution are summarized in Table~\ref{tab:runs}.

We can then estimate the detection probability as follows:
\begin{align}
    \alpha(\boldsymbol{\Lambda}) &= \int d\boldsymbol{\theta}p_{\mathrm{det}}(\boldsymbol{\theta})\pi_{\mathrm{pop}}(\boldsymbol{\theta} | \boldsymbol{\Lambda}),\\
    \alpha_{\mathrm{draw}} &= \int d\boldsymbol{\theta}p_{\mathrm{det}}(\boldsymbol{\theta})p_{\mathrm{draw}}(\boldsymbol{\theta}) \approx \frac{N_{\mathrm{found}}}{N_{\mathrm{draw}}}\\
    p_{\mathrm{found}}(\boldsymbol{\theta}) &= \frac{p_{\mathrm{draw}}(\boldsymbol{\theta})p_{\mathrm{det}}(\boldsymbol{\theta})}{\alpha_{\mathrm{draw}}},\\
      \alpha(\boldsymbol{\Lambda}) &= \alpha_{\mathrm{draw}}\int d\boldsymbol{\theta}\frac{p_{\mathrm{found}}(\boldsymbol{\theta})}{p_{\mathrm{draw}}(\boldsymbol{\theta})}\pi_{\mathrm{pop}}(\boldsymbol{\theta} | \boldsymbol{\Lambda}),\\
      \alpha(\boldsymbol{\Lambda}) &\approx \frac{1}{N_{\mathrm{found}}}\sum_{j} \frac{\pi_{\mathrm{pop}}(\boldsymbol{\theta_{j}} | \boldsymbol{\Lambda})}{p_{\mathrm{draw}}(\boldsymbol{\theta_{j}})}.
      \label{eq:alpha}
\end{align}
We account for the uncertainty in the Monte Carlo integral in Eq.~\ref{eq:alpha} following the method in \citet{Farr:2019rap} and reject parts of the hyperparameter space during sampling that do not have enough injections to meet the accuracy requirements therein. We note that our population model, $\pi_{\mathrm{pop}}(\boldsymbol{\theta_{j}} | \boldsymbol{\Lambda})$, in Eq.~\ref{eq:alpha} includes only the total mass distribution, and not the mass ratio distribution, as the latter has a negligible effect on the detectability of the source and would require a much higher number of injections to meet the accuracy requirements for our choice of narrow truncated Gaussian population distribution.

\bsp	%
\label{lastpage}
\end{document}